\documentclass[fleqn,usenatbib]{mnras}
\usepackage[T1]{fontenc}
\usepackage{ae,aecompl}
\usepackage{graphicx}	
\usepackage{amsmath}	
\usepackage{amssymb}	
\usepackage{pdflscape,xcolor,ae,aecompl,mathrsfs,soul,longtable}
\usepackage{hyperref}
\usepackage{natbib}
\usepackage{stfloats}
\usepackage{multirow}
\usepackage{color}
\usepackage{txfonts}
\usepackage{upgreek}
\usepackage{url}
\usepackage{epsfig}
\usepackage{epsf}
\usepackage{epstopdf}
\usepackage{float}
\usepackage{fancyhdr}
\usepackage{afterpage}
\usepackage{soul}
\usepackage{booktabs,caption}
\usepackage[flushleft]{threeparttable}
\usepackage[export]{adjustbox}
\usepackage{subfig}
\usepackage{nicefrac}
\usepackage{threeparttable, tablefootnote}

\newcommand{\laz}[1]{{\textcolor{black} {#1}}}
\newcommand{\adb}[1]{{\textcolor{black} {#1}}}


\newcommand{\planck}{\textit{Planck}}

\newcommand{\sigII}{\rm \sigma_{II}} 

\newcommand{\pa}{\psi}
\newcommand{\vdavgt}{VDA}   
\newcommand{\lowres}{$16\arcmin$}
\newcommand{\pobsmax}{$21.6$}
\newcommand{\pobsmean}{$4.9$}
\newcommand{\pobsmed}{$3.6$}

\newcommand{\pmaxref}{$21.6$}

\title[Polarization of the LMC with {\planck}]{Unveiling polarized emission from interstellar dust of the Large Magellanic Cloud with {\planck}}
{\author[D. Alina et al.]{
D. Alina$^{1,2}$\thanks{dana.alina@nu.edu.kz},
J.-Ph. Bernard$^{2}$,
K. H. Yuen$^{3}$,
A. Lazarian$^{3,4}$,
A. Hughes$^{2}$,
M. Iskakova$^{5}$,
A. Akimkhan$^{6}$, 
\newauthor
A. Mukanova$^{1}$
\\
$^{1}$Department of Physics, School of Sciences and Humanities, Nazarbayev University, Kabanbay batyr ave, 53, Nur-Sultan 010000, Kazakhstan\\
$^{2}$IRAP, Université de Toulouse CNRS, UPS, CNES, F-31400 Toulouse, France\\
$^{3}$Department of Physics, University of Wisconsin-Madison, Madison, WI 53706, USA\\
$^{4}$Centro de Investigación en Astronomía, Universidad Bernardo O’Higgins, Santiago, General Gana 1760, 8370993, Chile\\
$^{5}$Washington University in St Louis, One Brookings Drive, St. Louis, MO 63130\\
$^{6}$Nazarbayev Intellectual School, Khussein Ben Talal, 21, Nur-Sultan 010000, Kazakhstan
}}
\date{Accepted XXX. Received YYY; in original form ZZZ}

\pubyear{2022}
\hypersetup{colorlinks}

\begin{document}
\label{firstpage}
\maketitle

\begin{abstract}
Polarization of interstellar dust emission is a powerful probe of dust properties and magnetic field structure. Yet studies of external galaxies are hampered by foreground dust contribution.
The study aims at separating the polarised signal from the Large Magellanic Cloud (LMC) from that of the Milky Way (MW) to construct a wide-field, spatially complete map of dust polarization using the {\planck} 353 GHz data.
To estimate the foreground polarization direction, we used velocity gradients in HI spectral line data and assessed the performance of the output by comparing to starlight extinction polarization. 
We estimate the foreground intensity using the dust-to-gas correlation and the average intensity around the LMC and we assume the foreground polarization to be uniform and equal to the average of the MW around the galaxy to derive foreground I, Q, and U parameters.
After foreground removal, the geometry of the plane-of-the-sky magnetic field tends to follow the structure of the atomic gas. This is notably the case along the molecular ridges extending south and south-east of the 30 Doradus star-forming complex and along the more diffuse southern arm extending towards the Small Magellanic Cloud. 
There is also an alignment between the magnetic field and the outer arm in the western part.
The median polarization fraction in the LMC is slightly lower than that observed for the MW as well as the anti-correlation between the polarization angle dispersion function and the polarization fraction. 
Overall, polarization fraction distribution is similar to that observed in the MW.

\end{abstract}





\section{Introduction}

Along with turbulence and gravity, the magnetic field is thought to play a crucial role in star formation and in shaping the interstellar medium (ISM). It is therefore essential to consider magnetic fields in the studies of the ISM, and, first of all, to map the geometry of the magnetic fields in galaxies.
The Large Magellanic Cloud (LMC) is an irregular galaxy \citep{wielebinski1993} with low metallicity \citep{olszewski1991,Matsuura2009,grady2021}. It is the closest gas-rich galaxy and is viewed almost face-on. 
Polarimetric observations of interstellar dust reveal information about the magnetic field and the dust physics. The direction of the observed polarization is related to the plane-of-the-sky (POS) projection of the magnetic field. The amount of polarization depends in principle on dust properties and the intensity of the ambient radiation \citep{lazarian2007}, but for diffuse interstellar media it is  shaped primarily by the 3D structure of the magnetic field \citep{planck2014-xix}.

The alignment necessary to produce a net polarisation is thought to occur through phenomena such as Radiative torques (RATs) which lead to align the shortest grain axis parallel to the ambient magnetic field.
Consequently, the dichroic absorption of the incident starlight by these elongated grains yields the observed visible linear polarisation to be parallel to the POS magnetic field direction, while in emission, the polarization direction is perpendicular to POS magnetic field direction.

The LMC magnetic field was first studied using visible  extinction measurements in polarization \citep{visvanathan1966}. Spatially incomplete measurements showed that the LMC has an ordered large-scale magnetic field over some regions, and that the polarization is maximum in the bright 30 Doradus region \citep{visvanathan1966,mathewson1970,schmidt1976}.
More recently \citet{Wisniewski2007} mapped the optical polarization with imaging polarimetry around the NGC2100 supershell and showed regular polarization direction over the $15 \arcmin$ field of view. 
Near-infrared polarimetric observations complement optical observations which are limited to very low interstellar extinctions.
For instance, \citet{kim2016} examined a $\simeq 1$ square degree region in the Northern part of the LMC (in Galactic coordinates) in $J,H,K$ bands and inferred, using the DCF method (Davis-Chandrasekhar-Fermi, \cite{chandrasekhar1953,davis1951}),  a mean field strength of around $3-25 \, \mu G$ to be aligned with molecular clouds within that region.

At larger scales, synchrotron emission radio observations and Faraday rotation measures (RM) have been used to study the magnetic field of the LMC. \cite{haynes1990} showed that the polarization direction to be roughly aligned with filamentary structures seen in HI near the 30 Doradus region. \cite{gaensler2005} inferred the magnetic field strength to be larger than $1 \mu G$ and to have a spiral shape with a random component that dominates the ordered component by a factor of $\simeq$3, based on RM.
 \cite{mao2012} also analysed RM data and suggested the large-scale magnetic field to be spiral axisymmetric with the vertical component which reverses orientation across the disk and to be aligned with HI filaments. 
Still, it is likely that this structure applies to the large scale-height sampled by cosmic ray particles and does not probe the geometry of the field in the thin disk where most of the molecular material forming stars is located.

The large-scale structure of the magnetic field in the LMC associated with dense regions is still largely unexplored.
Optical and near-infrared observations of dust polarization in the LMC are sparse, and radio observations bring information about the line-of-sight (LOS) magnetic field component located mostly beyond the thin disk.
Here, we aim to complement the knowledge of the LMC magnetic field structure, taking advantage of the {\planck} satellite measurements at sub-millimeter wavelengths. 

Thermal dust in the LMC has been shown to have striking differences from what is observed in the Milky Way (MW). In particular, \cite{Bernard2008} has shown the presence of excess emission near $70\micron$ and in the FIR. This was confirmed by the analysis of the Planck data towards the LMC and the SMC in \cite{Planck2011_LMC_SMC}, which showed much flatter thermal dust submm spectral energy distributions (SEDs) in those galaxies than in the MW. Many low metalicity galaxies seem to share a similar behaviour \citep[e.g.,][]{RemyRuyer2013}. The origin of these differences are not clearly understood today. Some of the possible explanations invoke magnetic nanoparticles \cite[e.g.][]{DraineHensley2012} with specific predictions in polarization. It is therefore of interest to see if those differences also reflect in polarization, in order to gain knowledge on their possible origin.

The purpose of this study is to derive the dust polarization structure of the LMC using the {\planck} data.
To achieve this goal, a key challenge is to estimate and subtract the foreground polarization. 
The emission of the interstellar dust grains is maximum at sub-millimetre wavelengths where the radiation is optically thin \citep{draine2003}. 
This allows us to study distant objects, including those located outside of our Galaxy.
However, a particular attention should be paid to quantifying the contribution of the foreground material to the total radiation even on the LOS out of the Galactic disk, because the polarised emission of the interstellar dust is observed not only in the disk of the Galaxy, where the dust is mixed with the molecular gas, but also at high Galactic latitudes ($|\textit{b}_{II}| > 30^{\circ}$).
There, the emission of the diffuse dust and atomic gas, traced by far-infrared continuum and HI emission respectively, are observed to be correlated \citep{boulanger1996, kalberla2021}. 
Moreover, the polarisation fraction in the diffuse ISM is high compared to what is observed in dense regions: \cite{planck2014-xix} showed that the average polarization fraction at the column densities $10^{21}$ cm$^{-2}$ is around $7\%$ and can be up to $\simeq 20\%$ for lower column densities.
Thus, it is necessary to account for foreground contamination in polarization data. 

A first attempt to estimate foreground polarized extinction in the optical was made by \citet{visvanathan1966}. \citet{mathewson1970} compiled different star catalogs and provided catalogs of starlight polarization towards the LMC. It was shown that most of the foreground stars show polarization oriented in a similar direction, which was later confirmed by the study of \citet{schmidt1976}.
There are two main difficulties when estimating the contribution of foreground material to the observed dust polarisation using stellar extinction measurements. First, the point-like stellar extinction measurements must be robustly extrapolated across the entire field with dust polarisation information. Second, the target stars are located at various depths within the dusty material that is responsible for the extinction, and hence sample different columns of interstellar material.
In emission however, we would like to construct a fully sampled map of the foreground polarized emission, and that this map includes the contribution of all MW material along the LOS.
Therefore, we test two techniques to infer maps of the polarization direction of the foreground emission which we compare to the measured stellar polarization directions, and we use the observed polarized emission around the LMC as a proxy for the polarization fraction of the foreground towards the LMC.

In order to explore the foreground polarization structure in this paper, we use two recent  approaches that have established a link between the magnetic field direction and the properties of interstellar matter. The attractive feature of these approaches is that they use widely available
data for the atomic gas distribution and kinematics to infer the magnetic field structure. Both techniques use fluctuations of intensity in channel maps, but analyze them differently. 

The first technique is based on the notion of the magnetically aligned density filaments in cold HI data that can be detected using the Rolling Hough Transform algorithm \citep{clark2014}. The latter is an algorithm designed to detect linear structures in two-dimensional images. Using this approach, \citet{clark2019} constructed three-dimensional maps of the linear Stokes parameters using HI data and showed them to be statistically consistent with filaments in the {\planck} data, for which parallel alignment with the magnetic field in the diffuse ISM was previously reported \citep{planck2015-XXXV,alina2019}.

The second technique employs a prediction of MHD turbulence theory \citep{GS95,BL19} and the theory of turbulent reconnection \citep{LV99} that rotation axes of magnetic eddies are aligned with the local direction of the magnetic field \citep{LV99,cho2000}. Such aligned eddies induce the maximal variance of the velocity in the direction perpendicular to the eddy rotation axis. Therefore, the gradients of the velocity amplitude in MHD turbulence get maximal in the direction perpendicular to the local magnetic field, i.e. the gradient direction acts as a proxy of polarization. 
The corresponding Velocity Gradients Technique (VGT) has been applied not only to HI data, but also to CO, CHN and other molecular and atomic transitions, i.e. to the settings where the existence of filaments in real space has not been reported. These studies have shown a good correspondence between the magnetic field directions inferred using VGT and that obtained from polarimetry, e.g. with the {\planck}, BLASTPOL, and HAWC+ data  \citep{hu2019,alina2021,hu2020,hu2021,liu2022}.
For the VGT, the contribution arising from the actual HI density structures provides an unwelcome interference. To minimize its effects, we employ a recently developed Velocity Decomposition Algorithm (VDA) approach \citep{yuen2021} that separates the velocity and density fluctuations within the spectral data cubes.
The property of intensity gradients being related to the magnetic field has also been used in other studies, e.g., \cite{koch2012} to infer magnetic field strength in molecular clouds.

In this paper, we use both techniques to estimate the contribution of the Milky Way foreground to the observed polarization signal. We apply the techniques to HI data obtained by the GASS survey \citet{mcclure-griffiths2009}. In order to assess the accuracy of the magnetic field directions inferred using these techniques, we compare their results to polarization directions measured via optical polarization of stars by \citet{mathewson1970} and \citet{schmidt1976}. 

The paper is organized as follows. In Sect.~\ref{sec:data} we describe data sets used in the study. In Sect.~\ref{sec:methods} we briefly present the two existing techniques for estimating the magnetic field direction and describe our method for foreground removal. We present and discuss our results in Sect.~\ref{sec:results}. 
In Sect.~\ref{sec:summary} we summarize our key conclusions and discuss future perspectives.

\section{Data}
\label{sec:data}

\subsection{HI data}
\label{sec:HI_data}

\begin{figure}
    \centering
    \includegraphics[width = 0.5 \textwidth]{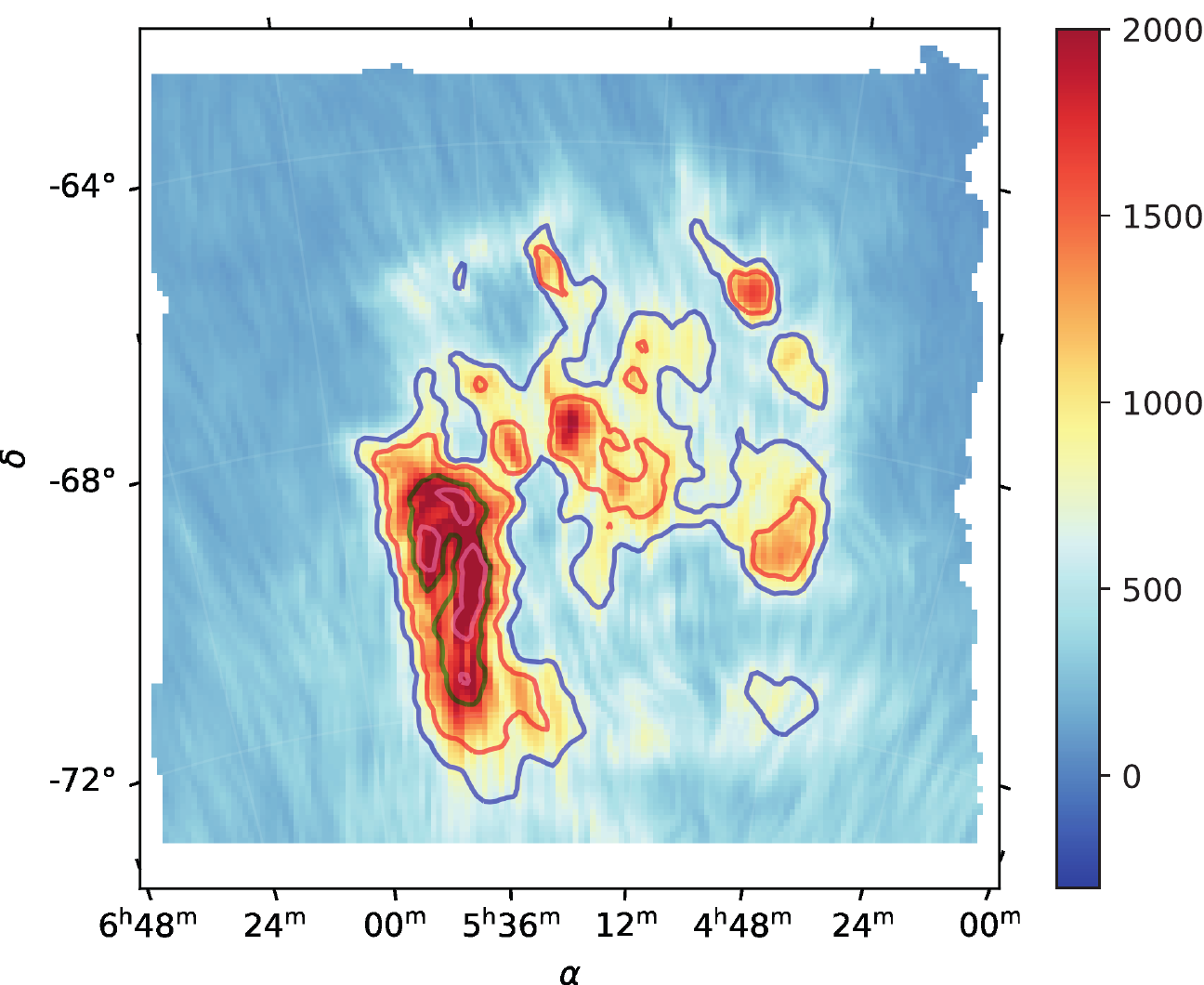}
    \caption{HI integrated intensity map, in K km s$^{-1}$, towards the LMC. It is overlaid with a drapery pattern showing the POS magnetic field direction derived from the {\planck} data at {\lowres}, computed using the LIC technique. The blue, red, green and pink contours correspond to the HI integrated intensity levels of $1000,1400,2400$, and $2800$ K km s$^{-1}$, respectively.}
    \label{fig:HItot}
\end{figure}

\begin{figure}
    \begin{center}
    \includegraphics[width = 0.48 \textwidth]{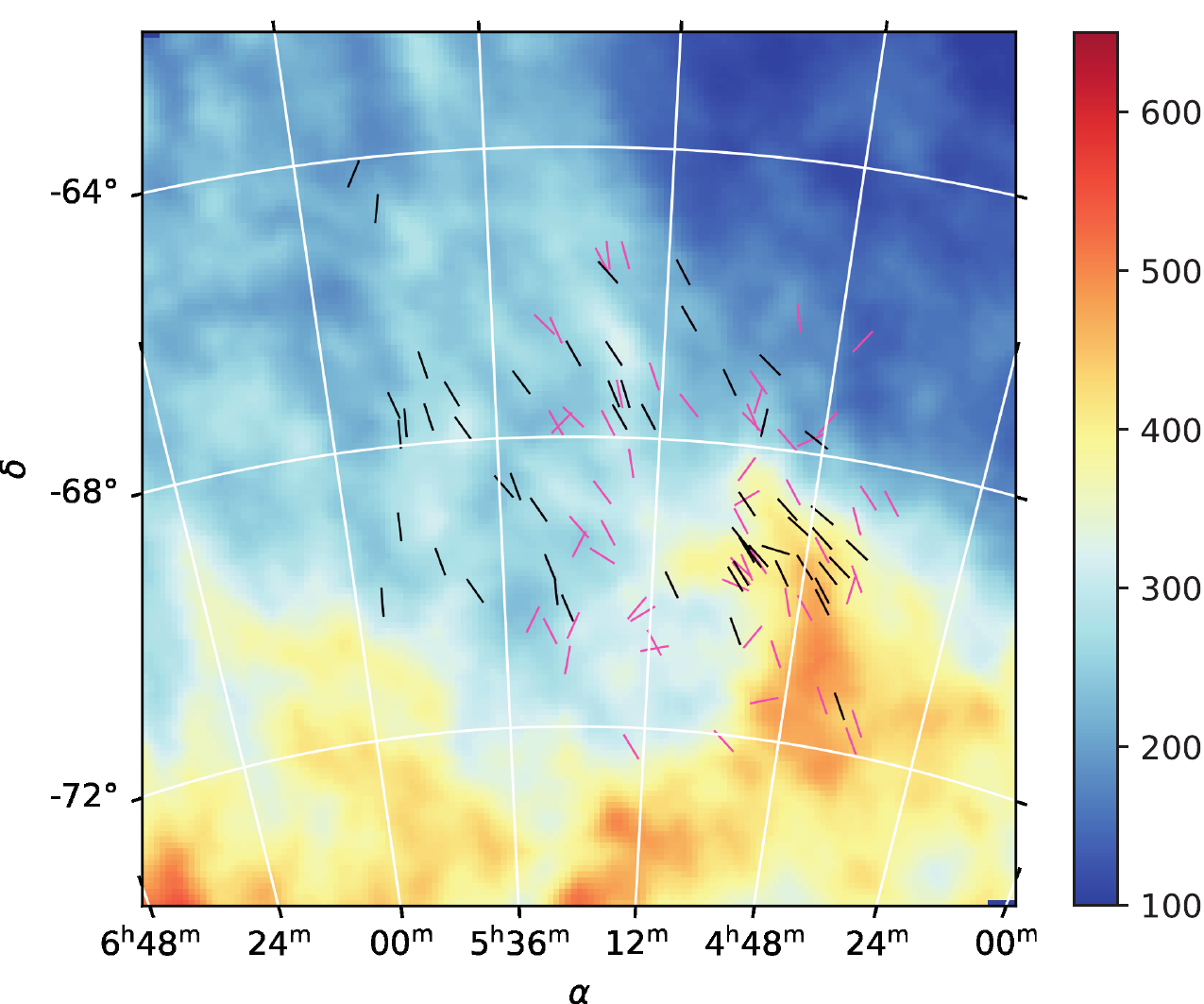}
    \end{center}
    \caption{Map of the HI foreground emission, in K km s$^{-1}$, integrated over $-50 < v < 90$ km s$^{-1}$. The segments show the magnetic field direction inferred from visible extinction (see Sect.\,\ref{sec:stars}) of the foreground stars. Pink segments show all available data entries, while black segments show the magnetic field directions for stars with distances larger than $500$ pc and with uncertainties on the polarization angle lower than $20^{\circ}$ used in this study.}
    \label{fig:HI_fg_lic_stars}
\end{figure}
We use the third data release of the Galactic All-Sky Survey (GASS), which observed the 21-cm HI emission with the Parkes 64-m Radio Telescope between January 2005 and October 2006 \citep{GASS2015,mcclure-griffiths2009}. This data release has been corrected for stray radiation and cleaned for radio frequency interference. The data has an effective spatial resolution of 16.2'. The survey covers local standard of rest (LSR) velocities ranging from -495 km/s to 495 km/s, with a channel separation of 0.82 km/s. The total integrated intensity map towards the LMC is shown in Figure~\ref{fig:HItot} with the magnetic field direction inferred from polarization in emission in the {\planck} data shown using the line integral convolution (LIC) technique \citep{cabral1993}.

To separate the foreground Galactic emission and the emission from the LMC, we use the velocity ranges proposed by \citet{staveley-smith2003} and use velocities from $-50$ km s$^{-1}$ to $90$ km s$^{-1}$ for the Milky Way, and from $182$ km s$^{-1}$ to $366$ km s$^{-1}$ for the LMC. Hereafter, the HI maps computed over the adapted Milky Way velocities range above will be referred to as the foreground maps. The foreground HI map is shown in Figure~\ref{fig:HI_fg_lic_stars}.

The total gas column density, used to provide the average column density further in Table~\ref{tab:p} was calculated from Parkes HI and NANTEN CO \citep{fukui2008} data, assuming optically thin HI and CO conversion factor of $\mathrm{X_{CO}}=2\,10^{20}$ cm$^{-2}$ K km s$^{-1}$.

\subsection{{\planck} data}
\label{sec:planck_data}

\begin{figure}
    \begin{center}
    \includegraphics[width = 0.5 \textwidth]{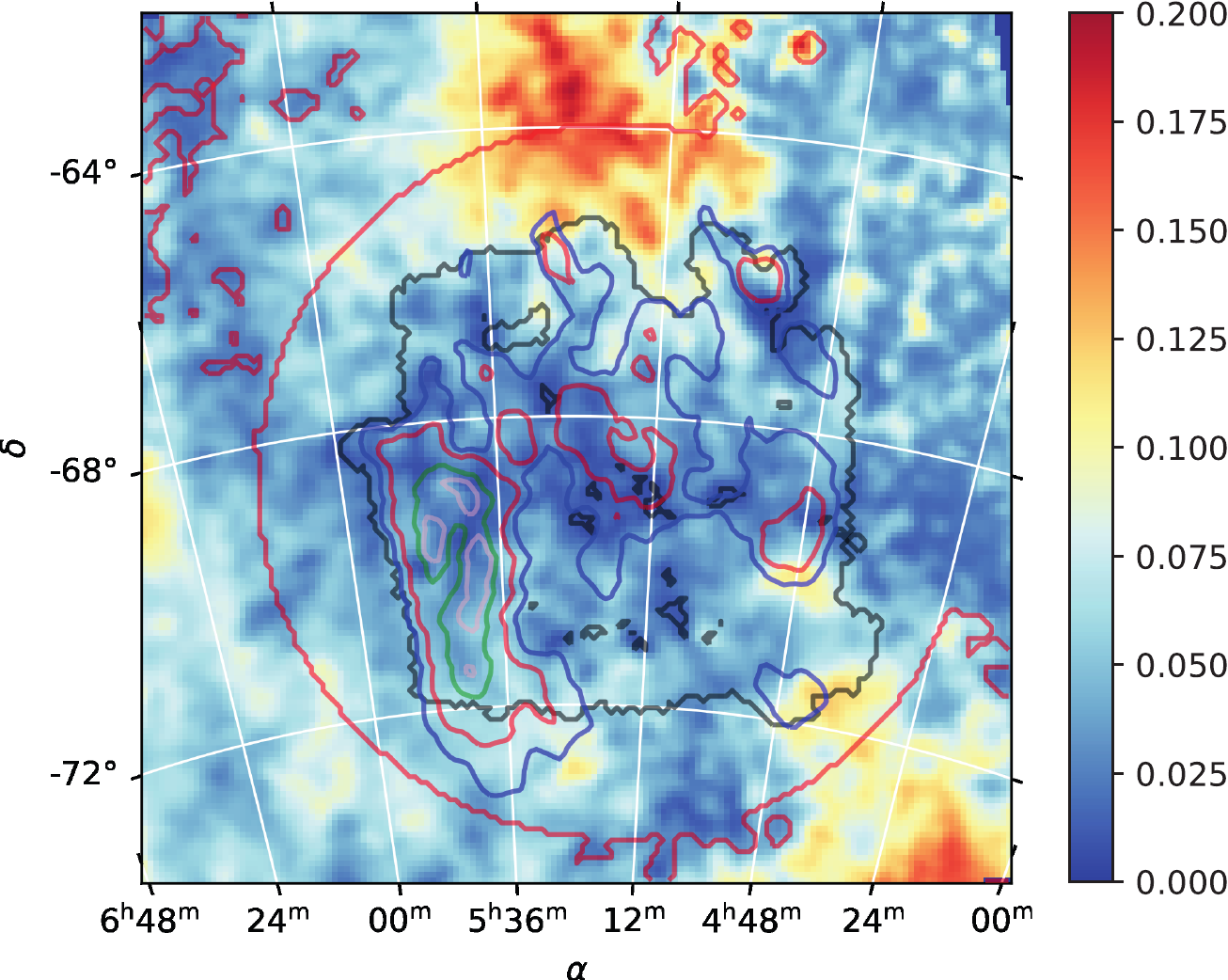}
    \caption{Map of polarization fraction derived from the {\planck} 353 GHz data, using the Bayesian approach (Sect.~\ref{sec:planck_data}) for the estimation of $p$. The black contour corresponds to the mask described in Sect.~\ref{sec:planck_data}. The contours are the same  HI integrated intensity levels as in Figure~\ref{fig:HItot}. The red contour shows the area excluded when computing the average foreground polarization fraction.}
    \label{fig:pmap}
    \end{center}
\end{figure}

The {\planck}\footnote{{\planck} (\url{http://www.esa.int/planck}) is an ESA science mission designed and completed by a collaboration of institutes directly funded by ESA Member States, NASA, and Canada.} instrument mapped the whole sky over a range of frequencies from 70 to 857 GHz to measure the Cosmic Microwave Background (CMB) anisotropies and Galactic foregrounds.
Here, we use the {\planck} PR3\footnote{The {\planck} data is available at \url{https://pla.esac.esa.int/\#home}} data release. The CMB contribution is subtracted using the Generalized Needlet Internal Linear Combination algorithm \citet{remazeilles2011,planck2018-iv}.   
We also subtracted the Cosmic Infrared Background (CIB) monopole contribution to the intensity map according to \citet{planck2013-XI}.
For polarization, we restrict the analysis to the $353$ GHz channel, which is the highest frequency channel of {\planck} with polarization capabilities.

The LMC was observed by {\planck} many times, since the {\planck} scanning strategy \citep{planck2011-1.1} repeatedly covered the region close to the ecliptic poles. In particular, the LMC is located at the boundary of the {\planck} deep field. Thus, the variation of the noise level across the LMC is one order of magnitude in intensity, as illustrated in the left panel of Figure~\ref{fig:snrI}, and the signal-to-noise ratio (SNR) of the intensity within the LMC is $\sim$1000 (see right panel of Figure~\ref{fig:snrI}). Given that these observations were repeated with many different analysis angles, the {\planck} data in polarization at 353 GHz in the direction of the LMC are in principle extremely reliable.

The polarization fraction $p$ and angle $\pa$ can be derived from the  Stokes parameters $I,\, Q,$ and $U$ observed by {\planck} according to following equations:
\begin{equation}
p = \frac{\sqrt{Q^2+U^2}}{I} \, ,
\label{eq:polfrac}
\end{equation}
\begin{equation}
\pa = 0.5 \mathrm{atan} (-U, Q) \,.
\label{eq:polang}
\end{equation}
The negative sign in front of $U$ is to pass from the COSMO (or Healpix) convention used for the {\planck} data to the IAU convention for polarization. 
$p$ and $\pa$ are biased non-linearly due to noise propagation in equations \ref{eq:polfrac} and \ref{eq:polang} \citep{serkowski1958,simmons1985,vaillancourt2006,quinn2012,Montier1}. Estimates of $p$ and $\pa$ must be corrected for this bias.

We compute the Mean Posterior estimates of $p$ and $\pa$ using the Bayesian method described in \cite{Montier2}, which takes into account the noise variance matrix of each map pixel. To do so, we performed Monte-Carlo simulations with $10^7$ realizations. The advantage of the Bayesian method is that it allows us to compute the polarization angle and polarization fraction, and their uncertainties from a common probability distribution function (PDF), even at low SNR (SNR($p$) $\simeq 2$). The resulting polarization fraction is shown in Figure~\ref{fig:pmap}.
Other existing methods only provide estimates of the polarization fraction or angle that are valid at intermediate to high SNR, SNR($p$) $> 3$ \citep{simmons1985,Montier1}. 

We brought the {\planck} data to the resolution of the HI data. To do so, we degraded the HEALPix\footnote{\url{https://healpix.sourceforge.io/}} {\planck} maps at nominal angular resolution ($\simeq 5\arcmin$) to {\lowres} and pixel size corresponding to $N_{side} = 1024$, according to the procedure described in \cite{planck2014-xix}. 
This method takes into account the rotation of the local polarization reference frame in the convolution. 
The angular size of {\lowres} at the distance of the LMC corresponds to a physical size of 200 pc.

The local maps used in this study were computed using the Drizzlib\footnote{\url{http://cade.irap.omp.eu/dokuwiki/doku.php?id=drizzlib}} approach, which is designed to calculate the intersection of pixel surfaces precisely and hence preserves the photometry accuracy.

In this paper, we only derive polarization parameters of the LMC for the data that satisfy the following criteria: the original intensity is larger than the median value of $I_{med} = 0.276$ MJy/sr inside the region shown in the map in Figure~\ref{fig:pmap} and the signal-to-noise ratio of polarization fraction is larger than three, $p/\sigma_p > 3$, where $\sigma_p$ is computed using the Bayesian method described above. The resulting mask is shown as the black contour in Figure~\ref{fig:pmap}.

\subsection{Optical polarization data}
\label{sec:stars}

\begin{figure}
    \begin{center}
    \includegraphics[width = 0.49 \textwidth]{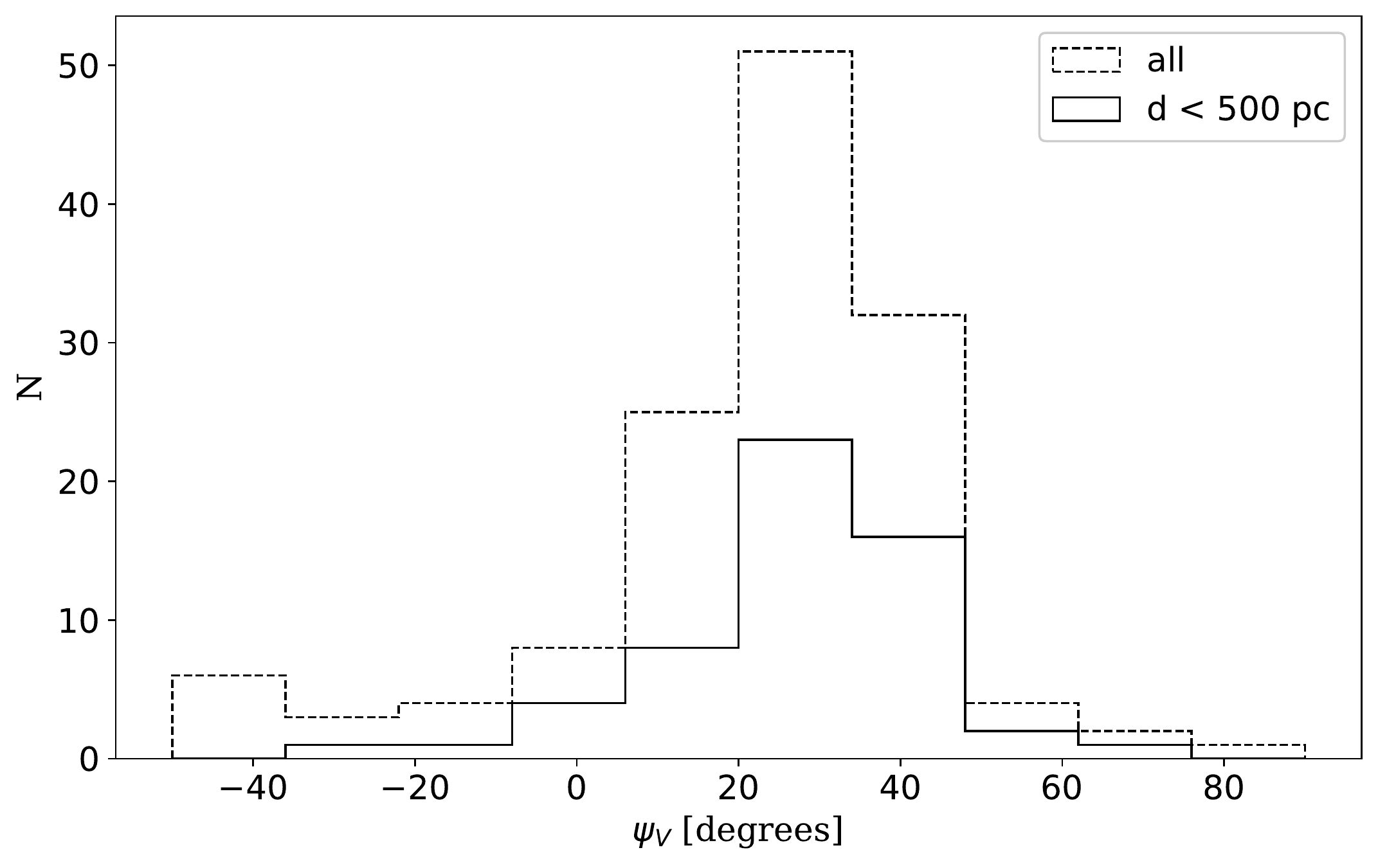}
    \caption{Histograms of the visible polarization angles towards foreground stars. The polarization angle is measured from the local North direction and is gven in the IAU convention. The dashed line shows all available data entries, while the solid line shows the data for stars with distances larger than $500$ pc and with uncertainties on polarization angle lower than $20^{\circ}$ used in this study. 
    }
    \label{fig:stars_hist}
    \end{center}
\end{figure}

The optical polarization data used are taken from following catalogues:
\renewcommand{\labelitemi}{$-$}
\begin{itemize}
    \item The \cite{schmidt1976} (hereafter S76) catalogue contains optical polarization data towards the Magellanic Clouds (MC) region. The polarization observations for 309 stars were obtained between 1971 and 1972, using the ESO 1m photometric telescope and the ESO two channel polarimeter at La Silla observatory in Chile. The catalogue also contains data from \cite{schmidt1970} obtained from the Boyden observatory. In total for the LMC, there are 149 galactic foreground stars. We manually copied the polarization data for foreground stars from Table 6 of \cite{schmidt1976} original paper. 
    \item \cite{mathewson1970} (hereafter MF70) catalogue contains optical polarization measurements taken during 1969 and 1970 using the 74-inch reflector at Mount Stromlo and the 24-inch telescope at Siding Spring Observatory. There are 31 foreground galactic stars toward the LMC.  
\end{itemize}
The polarization angles in the above catalogues were originally given with respect to the equatorial coordinate system at epoch B1950, which we precessed to J2000.

In our star catalogs, distance estimates were missing or doubtful for a significant number of stars. As it is critical to control how far the Milky Way ISM is traced by the observations, we updated the distances and completed missing data using the SIMBAD database \citep{wenger2000}. The new distances were taken from the following Gaia-based \citep{gaia2016} catalogues: the Gaia 2nd release \citep{gaia2018,bailer-jones2018} and the StarHorse catalog of astro-photometric distances \citep{anders2020}. 
In total, we compiled distance data for 145 of the 149 stars in the S76 catalog, and 30 of the 31 stars in MF70 catalog. In case of discrepancies between the original polarization catalogues and recent distance estimated, we adopted the latter. This represents 97.3$\%$ and 96.8$\%$ of entries in the S76 and MF70 catalogs respectively.
The compiled data is publicly available upon publication\footnote{\url{https://github.com/danakz/starpol}}.
We used background stars for which the distance is larger than $500$ pc in order to ensure that the corresponding polarization traces the foreground magnetic field direction over most of the corresponding emitting LOS. We also retained stars for which the uncertainty on the polarization angle is less than $20^{\circ}$. These criteria yield a final sample of 56 foreground stars. 

The magnetic field position angles inferred from the stellar extinction data are shown in Figure~\ref{fig:HI_fg_lic_stars}. 
The foreground magnetic field directions appear to be rather uniform, oriented almost vertically and close to the North-East to South-West direction.  In Figure~\ref{fig:stars_hist}, we present histograms for all stars and for our final selection. We observe that restricting the selection to star distances larger than $500$ pc does not significantly change the pattern. However, the corresponding histogram is narrower.
The direction of the visible polarization is globally aligned parallel to the magnetic field structure around the LMC, traced by {\planck}.
Also, the HI background map shows a clear elongated structure crossing the LMC. This structure dominates the overall foreground, and actually extends further than shown in Figure~\ref{fig:HI_fg_lic_stars}, with a similar orientation. The magnetic field direction inferred from our stellar data appears to globally follow that direction, which is consistent with the expectation of a magnetic field oriented parallel to density filaments \citep{clark2019}. This motivates using filaments identification to constrain the magnetic field direction. 

\section{Methods}
\label{sec:methods}
The separation between the polarization originating from the LMC and that from the MW foreground is best achieved through the Stokes parameters formalism.
Given that both the MW foreground and the LMC emission is optically thin, we assume 
\begin{eqnarray}
    I_{LMC} &=& I_{obs} - I_{fg} \\
    Q_{LMC} &=& Q_{obs} - Q_{fg} \\
    U_{LMC} &=& U_{obs} - U_{fg}\, .
    \label{eq:global}
\end{eqnarray}
where $I_{obs}$, $Q_{obs}$ and $U_{obs}$ are the total Stokes parameters observed with {\planck} and the corresponding quantities with subscripts $_{LMC}$ and $_{fg}$ are for the LMC and MW foreground respectively.
The foreground Stokes parameters can be written as following:
\begin{eqnarray}
    Q_{fg} = p_{fg} \, I_{fg} \, \cos{2 \pa_{fg}}\label{eq:qu1} \\
    U_{fg} = p_{fg} \, I_{fg} \, \sin{2 \pa_{fg}}\label{eq:qu2} \,.
\end{eqnarray}
where $I_{fg}$ is the foreground total intensity and $p_{fg}$ and $\pa_{fg}$ are the assumed polarization fraction and angle of the foreground contribution respectively, which we estimate using the approaches described below.

\subsection{Determination of the foreground polarization angle}

To construct a map of the foreground polarization angle, we use two independent techniques that assume a relationship between the magnetic field direction and the intensity structures identified using spectroscopic data. The first uses the Rolling Hough Transform (RHT) to derive the direction of  magnetically-aligned filaments. The second uses the Velocity Gradient Technique (VGT) to constrain the field direction based on MHD assumptions. We apply the methods to the HI data that accounts for the foreground contribution, as described in Sect.~\ref{sec:HI_data}, that is from $-50$ to $90$ km s$^{-1}$.
We assess the performance of each method to this study by comparing their predictions with the magnetic field direction derived from optical starlight polarization data, assumed to be a robust measurement of the projected field orientation on those LOS where stellar data is available.

\subsubsection{The Rolling Hough Transform}
\begin{figure*}
    \begin{center}
    \begin{tabular}{cc}
     \includegraphics[width = 0.495\textwidth]{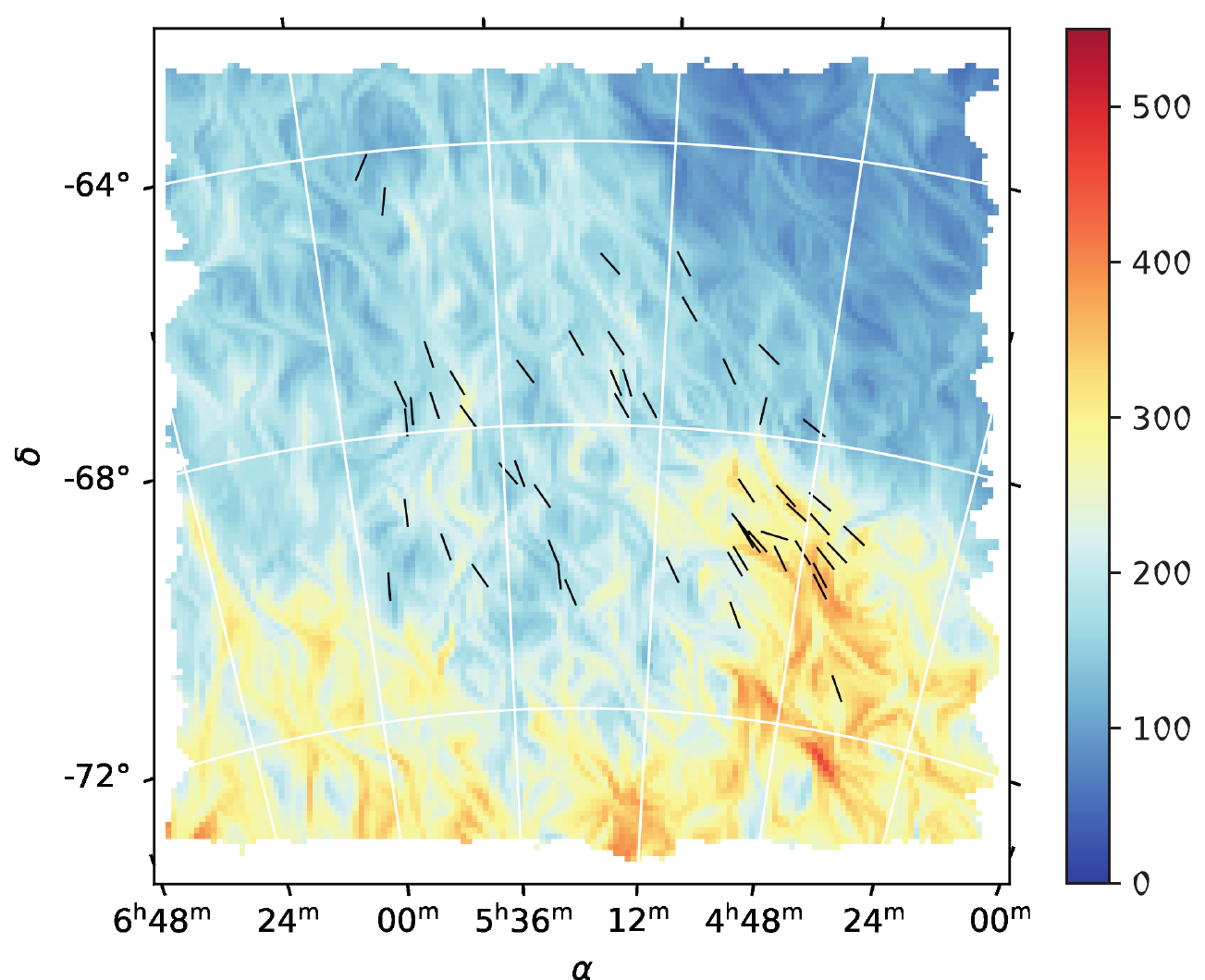}
     & 
     \includegraphics[width = 0.485 \textwidth]{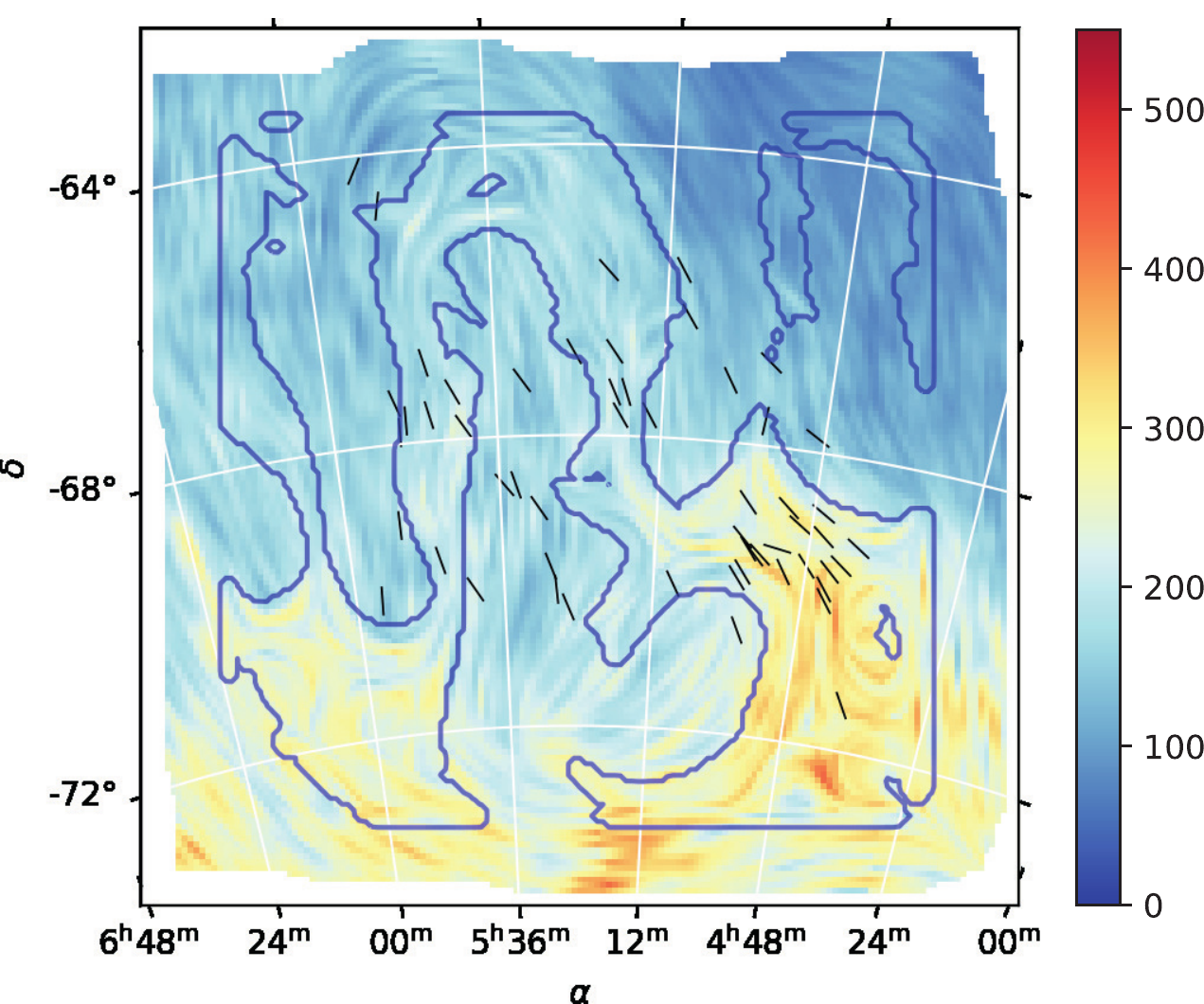}
     \end{tabular}
    \caption{\textit{Left panel}: Orientation of the RHT filaments computed by \citet{clark2019} represented by the drapery pattern produced using the LIC technique. \textit{Right panel}: Orientation of the RHT filaments detected in foreground HI map shown as blue segments. The blue contours represent the detected filaments, while the LIC-drapery pattern represents the interpolated angles.  The background image in both panels is the foreground HI map in K km s$^{-1}$, and the black segments show the direction of the magnetic field from extinction polarization towards foreground stars.}
    \label{fig:RHT_interp}
    \end{center}
\end{figure*}

\begin{figure}
    \begin{center}
    \begin{tabular}{cc}
    \hspace{-0.3 cm} \includegraphics[width = 0.24 \textwidth]{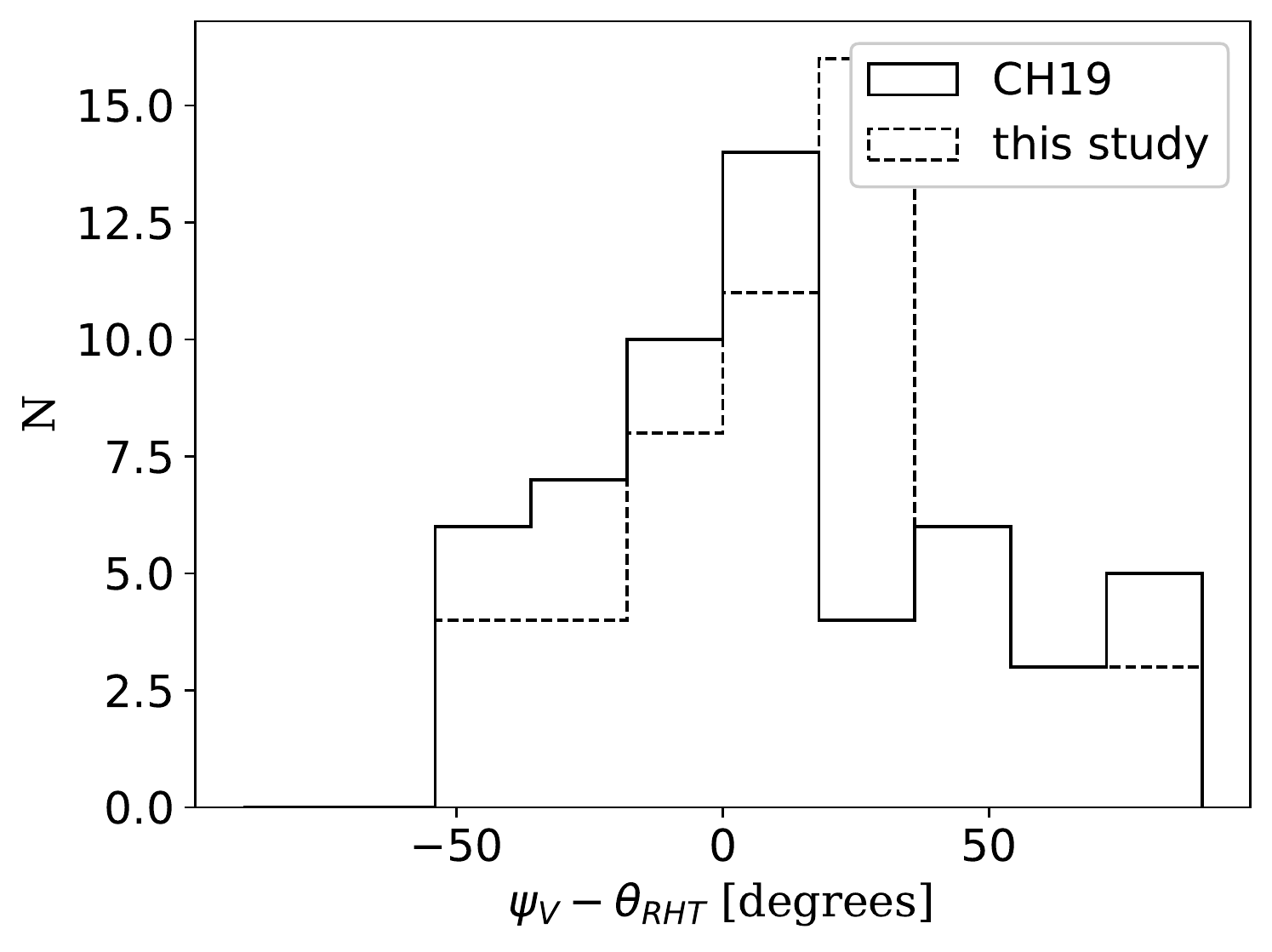} 
    & \hspace{-0.5 cm}
    \includegraphics[width = 0.24 \textwidth]{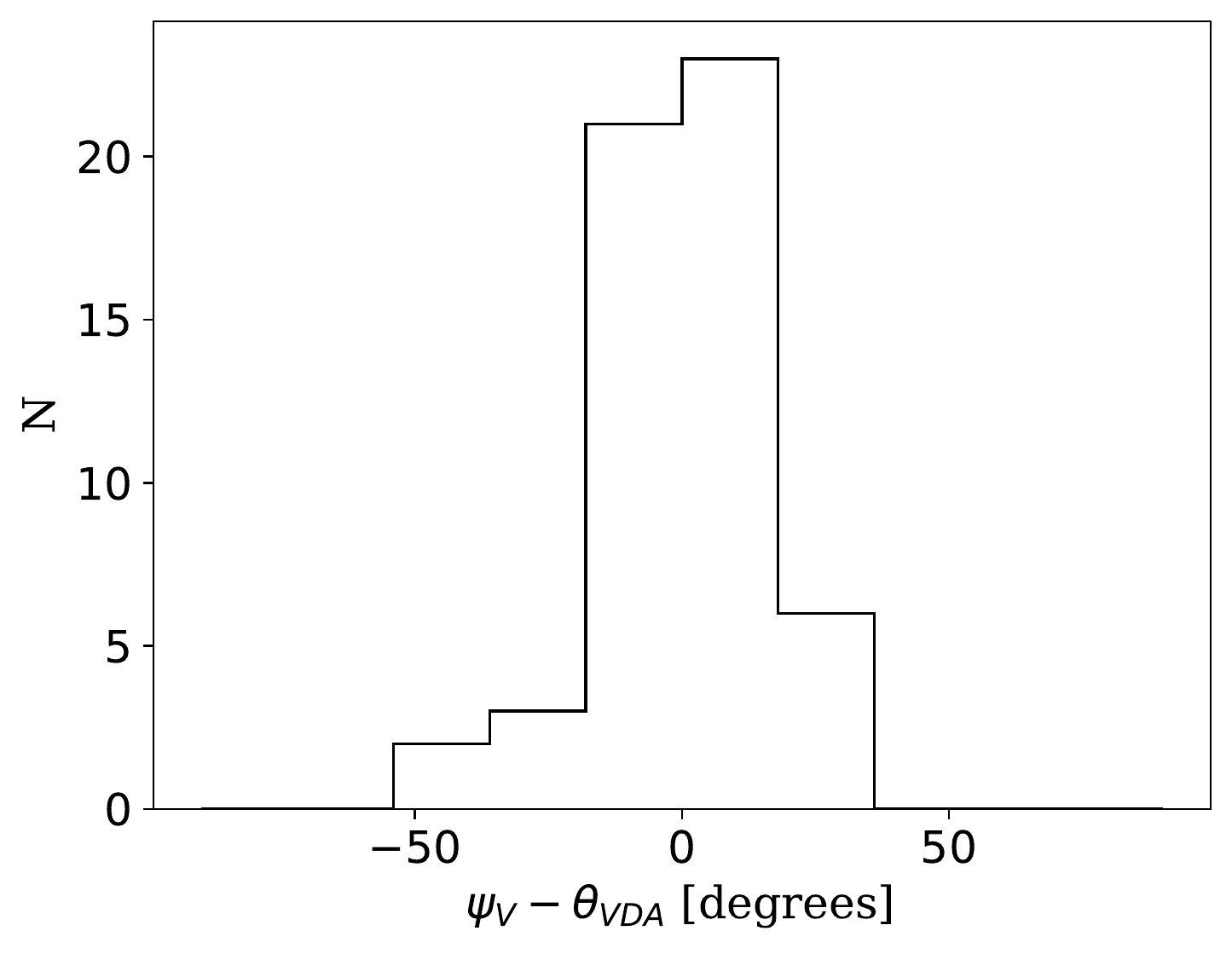}
    \end{tabular}
    \caption{\textit{Left panel:} histogram of the differences between the starlight extinction polarization angles and the orientation of the structures detected using RHT technique. The solid line corresponds to the orientations taken from \citet{clark2019} (CH19, filaments detected in thin velocity channels and then integrated over the velocities). The dashed line corresponds to the structures found using the same method applied to the integrated foreground HI map. \textit{Right panel}: histogram of differences between the starlight extinction polarization angles and the magnetic field direction of the foreground derived using the VGT technique.}
    \label{fig:RHT_angdiff}
    \end{center}
\end{figure}

\noindent Comparisons between the atomic hydrogen emission at 21 cm and magnetic field directions derived from polarization have shown that the diffuse HI gas is organized into elongated linear fibers, the orientation of which correlates with the orientation of the magnetic field \citep{clark2014}. Based on this observation, \citet{clark2019} suggested that it is justified to derive the magnetic field direction via the determination of the orientation of HI fibers.  While a detailed description of the RHT can be found in the original article, here we briefly describe the main steps involved in the method.

RHT is a machine vision algorithm that identifies and quantifies linear structures in two-dimensional maps. 
The input two-dimensional image is smoothed using a top-hat function of a given diameter defined by the user. The smoothed image is then subtracted from the original map and a threshold is applied to the difference map to obtain a bitmask.
Next, a disk of a given diameter is extracted around each point in the bitmask and the Hough Transform is applied to each disk. The Hough Transform measures the intensity of each rectangular kernel centered at the disk and stores the one above set threshold as $R(\theta, x_0,y_0)$, RHT intensity, as a function of angle for each pixel.
The RHT algorithm takes as input parameters a map, over which to operate, the size and width of the kernel window,
smoothing radius, and the intensity threshold as a percentage and returns a map of identified structures, their position angle values ($\theta_\text{RHT}$) and associated uncertainties ($\sigma_{\theta,\text{RHT}}$). One of the main advantages of this technique is that it determines  the linearity of structures that do not have sharp boundaries.
A disadvantage of the method is that the size and the shape of the identified structures strongly depends on the user-defined kernel parameters. 

\citet{clark2019} provided a full sky data set\footnote{https://dataverse.harvard.edu/dataverse/ClarkHensley} of $I, \, Q$ and $U$ in 3D (the third dimension being the velocity) of the filamentary structures identified using the RHT method applied to HI data from the GALFA \citep{peek2018} and the HI4PI \citep{benbekhti2016} surveys over the Galactic velocity range. We extract the data from the integrated RHT $Q$ and $U$ maps towards the LMC region. The left panel of Figure~\ref{fig:RHT_interp} shows the orientations of the structures derived from the RHT method as a LIC-drapery  and the starlight polarization data with the black segments. 
Although those orientations seem to globally match structures that are observed in the HI integrated intensity map, the directions of the detected filaments vary a lot. 
The RHT filaments directions do not correspond to the starlight polarization data very well as can be seen from the left panel of Figure~\ref{fig:RHT_angdiff} where a histogram of the angle difference is represented with black solid line.
This may be due to some of the following reasons. First, the RHT parameters are adjustable and those used in \cite{clark2019} can be sub-optimal for the region. In fact, the size of the kernel was fixed to $75'$ for the calculation for the whole Galaxy. Second, in the original work, the RHT procedure was applied in velocity channel bins, with the bin size fixed to $\Delta v = 3.7$km s$^{-1}$. The integration of the RHT Stokes $Q$ and $U$ over all the velocity channels, including channels with no obvious structures or less significant small features could yield many various directions to appear.

Even though \citet{clark2019} found a good correlation between the orientations of RHT-derived structures and the magnetic field direction derived from the {\planck} data across the ISM in our Galaxy, the corresponding maps do not seem to us to be the most appropriate in the direction towards the LMC.

We tried to improve the HI-filaments output to correspond better to the observed polarization and to address possible issues regarding the variation of the direction of filaments detected in different velocity-channels and the choice of the RHT parameters. To do so, instead of detecting filaments in thin velocity channels and then integrate over the velocities, we ran the RHT method over the integrated foreground HI map. The kernel in the RHT technique is an input parameter,
and we tried various kernel sizes. The best detection of HI intensity filaments that reflected visual perception of the HI map was achieved for the kernel with the length of $2^{\circ}$ and the width of $25\arcmin$. It is worth noting that we did not introduce any matching of the filaments with  the polarization data in our optimization of the RHT parameters.
Since filaments are only detected in the brightest regions, there are parts on the map that have missing values. We therefore interpolated the results via the Stokes parameters $Q_{RHT} = \cos(2\theta_{RHT})$ and $U_{RHT} = \sin{2\theta_{RHT}}$ using a spatial exponential interpolation:
\begin{equation}
    X'_{i}  = \frac{\sum\limits_{j}^{N} X_{RHT,j} \mathrm{e}^{-\frac{d_{ij}^2}{\sigma_d^2}}}{\sum\limits_{j}^{N}  \mathrm{e}^{-\frac{d_{ij}^2}{\sigma_d^2}}}\, ,
\end{equation}
where $X'_{i}$ is the interpolated $Q$ or $U$ parameter, $d_{ij}$ are the distances from a given pixel $j$ to pixels where RHT filaments were detected, $X_{RHT,j}$ with $j \in [1,N] $ are the Stokes parameters of these filaments and $\sigma_d$ is a characteristic length, taken to be $100\arcmin$, N is the number of pixels. 
We also smoothed the integrated foreground HI map at 20 pixels scale prior to the application of the RHT kernel. This gives the same spatial averaging as the one required in VGT calculation shown in the next Section and allowed us to compare the results.
The right panel of Figure~\ref{fig:RHT_interp} shows the directions of the derived filaments as well as the interpolated directions as a LIC-drapery pattern and the blue contours show the extent of the detected filaments.
The resulting map of RHT-directions using this approach seem to better follow the orientation of the observed elongated structures in the foreground HI map. However, the histogram of the difference with respect to the starlight polarization in the left panel of Figure~\ref{fig:RHT_angdiff} shows that the peak is shifted to 25 degrees. 
Also, the distance between the detected filaments in some cases is as large as the largest filament structure, and the spatial interpolation may be doubtful. Thus, although there is some parallel alignment between filaments and starlight polarization, there are also large differences for both RHT-based results.

\subsubsection{Velocity gradients technique}
\label{sec:vgt}

\begin{figure}
    \begin{center}
    \includegraphics[width = 0.5 \textwidth]{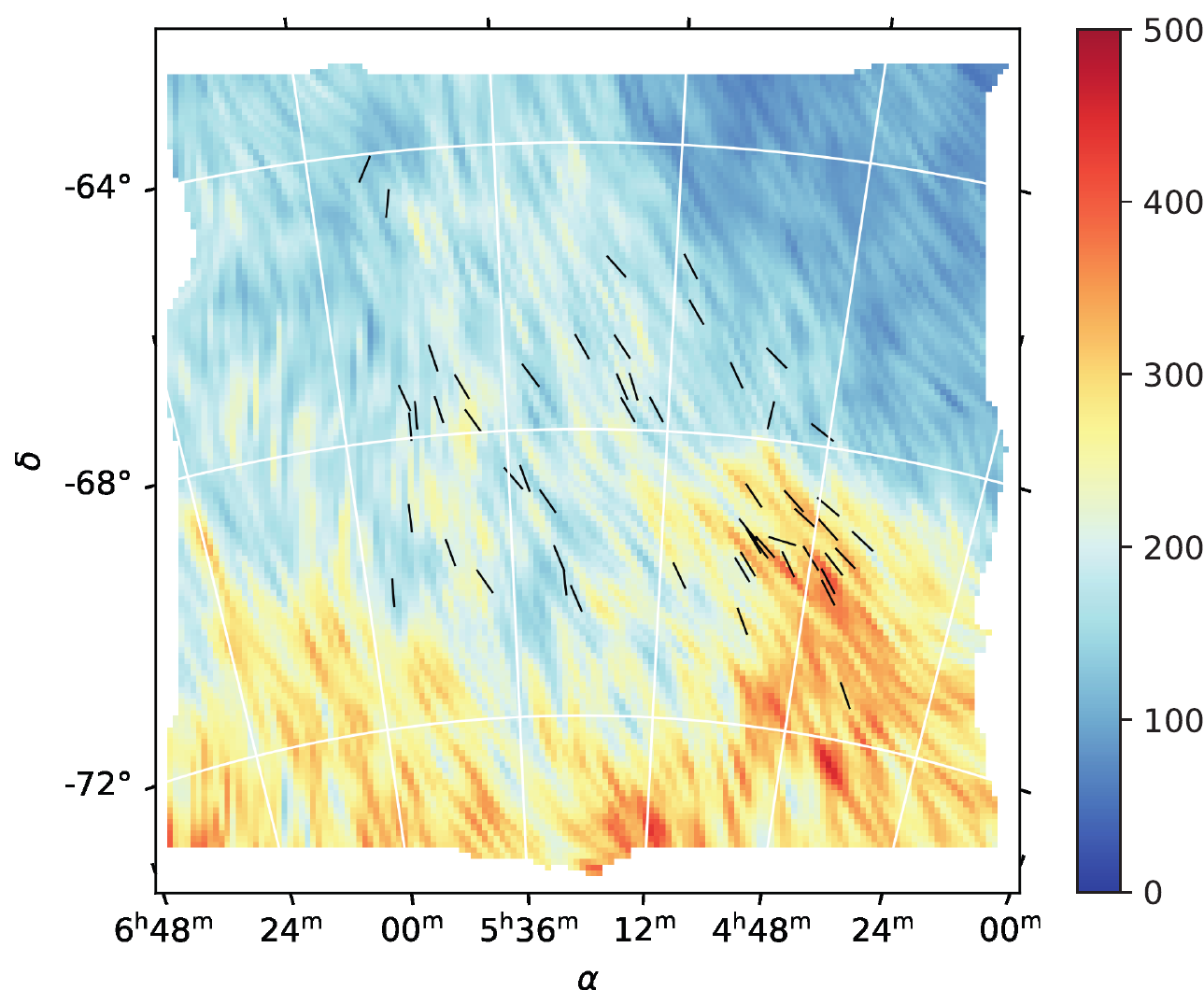} 
    \caption{Direction of the foreground magnetic field orientation obtained through the {\vdavgt} method, represented as the drapery pattern produced using the LIC technique. The background image shows the foreground HI map in K km s$^{-1}$. The black segments show the direction of extinction polarization of foreground stars.}
    \label{fig:vda_map}
    \end{center}
\end{figure}

\noindent Interstellar gas is turbulent \citep{armstrong1995,chepurnov2010} and the theory of 
magnetohydrodynamic (MHD) turbulence \citep{BL19} connects the turbulent motion of the plasma to the magnetic field structure. On this basis, the velocity gradients technique (VGT) was developed specifically to trace interstellar magnetic fields \citep{gonzales-casanova2017,yuen2017vgt,lazarian2018vgt}. 

The foundations to the technique are rooted in the turbulence anisotropy discovered in \citet{goldreich1995}.  Turbulent reconnection enables mixing of the magnetic field lines by eddies that are elongated along the local direction of the magnetic field \citep{LV99}.  Within these eddies, magnetic field lines reconnect and do not prevent eddies to evolve in a hydrodynamic fashion.  Since the eddies are closely aligned with the direction of the magnetic field that they are embedded in, the changes of amplitudes of velocities are maximal in the direction perpendicular to the magnetic field. Therefore the gradients of the velocity amplitude are expected to be perpendicular to the magnetic field direction and can trace the magnetic field orientation, in a way similar to magnetic field orientation being traced by the polarization of the dust emission (see also a pictorial illustration in \citet{pattle2019}). 

The velocity fluctuations are reflected in the  fluctuations of intensity in spectroscopic data cubes and the latter are most pronounced in thin channel maps \citep{LP00}. The analysis of gradients of intensity in channel maps are suggested to use in the version of the Velocity Gradient Technique (VGT) that we employ in the paper \citep{lazarian2018vgt}. It is clear from the description that while the RHT is focused on structures arising from density variations, the VGT is focused on exploring the effects induced by velocity fluctuations.    

It is worth noting that additional effects can influence velocity gradients. The propagation of shocks and the effect of gravitational collapse were considered in the literature as possible influencing factors. Shocks are most important for gradients of intensities \citep{yuen2017vgt,hu2019}. For velocity gradients, the effects of shocks is subdominant as the change in velocity is very localized and the sub-block averaging employed in the technique, and described further in this Section, suppresses the effect of such localized velocity amplitude changes. The velocity gradients arising from gravity can be important for localized regions where gravity-induced acceleration is larger than that induced by turbulence \citep{hu2019} In practical terms, these are relatively small collapsing regions of molecular clouds and are not relevant for the present study. 

The effects of gravity and shocks can be amplified if the employed measures of velocity are contaminated by density effect. This is more pronounced when using velocity centroids \citep{esquivel2005}, rather than when using individual channel maps \citep{LP00}. \laz{Thus, in the present paper we use the version of the VGT that employs velocity channels \citep{lazarian2018vgt} rather than maps of velocity centroids} \citep{gonzales-casanova2017,yuen2017vgt}.

More recently, on the basis of the theory in \citet{LP00}, \citet{yuen2021} proposed a new technique, the Velocity Decomposition Algorithm (VDA), for filtering effects of density in spectroscopic data. This approach significantly reduces the effects of density fluctuations and was shown to reduced the area over which the measured gradients of intensities in velocity channels are not perpendicular to the magnetic field orientation. In our study we employ the VDA technique to increase the accuracy of the VGT method. 

In practice, gradients are calculated in velocity-centroid or velocity channel maps for every pixel. Then, a histogram of the gradients is constructed for a sub-block of a given size. The size of the sub-blocks depends on the noise in the data and is chosen to get a distribution of gradient directions close to Gaussian. In this study, the conventional size of 20$\times$20 pixels \citep{yuen2017vgt} is employed, which corresponds to a total area of around 50 beams. The expectation value of the distribution gives the mean orientation of the gradients within a sub-block \citep{yuen2017vgt}, while the width of the distribution provides a measure of media magnetization, i.e. its Alfven Mach number, $M_A=V_L/V_A$, where $V_L$ is the turbulent injection speed and $V_A$ is the Alfven speed (Lazarian et al. 2018). In this paper, we  focus on the magnetic field directions derived from velocity gradients and do not study the $M_A$ distribution. 

The fitting of a Gaussian into the distribution of gradient directions provides a practical advantage of the VGT, i.e. the size of the sub-block is being determined on the basis of the objective criterion. The obtained VGT maps do not require extrapolation either. 

Our predicting of the galactic foreground polarization is similar to that in \cite{lu2020}, where the motivation was to improve the accuracy of distilling the Cosmic Microwave Background (CMB) polarization. The difference is that we use the VDA to remove the effects of densities and thus increase the accuracy of magnetic field tracing.  

The magnetic field orientation of the foreground as derived using by {\vdavgt} method is shown in Figure~\ref{fig:vda_map}. 
The orientation is globally in the North-East to South-West direction, close to the vertical, and the starlight extinction polarization is generally aligned parallel to this orientation. As can be seen from the right panel of Figure~\ref{fig:RHT_angdiff}, the histogram of the angles difference between the predicted magnetic field direction and the observed polarization in the visible is centered on zero, indicating a good match between the predictions of the method and the observed POS magnetic field orientation. In the following, we use the result from the VDA method to constrain the direction of the foreground polarization.

\subsection{Determination of the foreground intensity}
\label{sec:intensity}

\begin{figure}
    \begin{center}
    \includegraphics[width = 0.5 \textwidth]{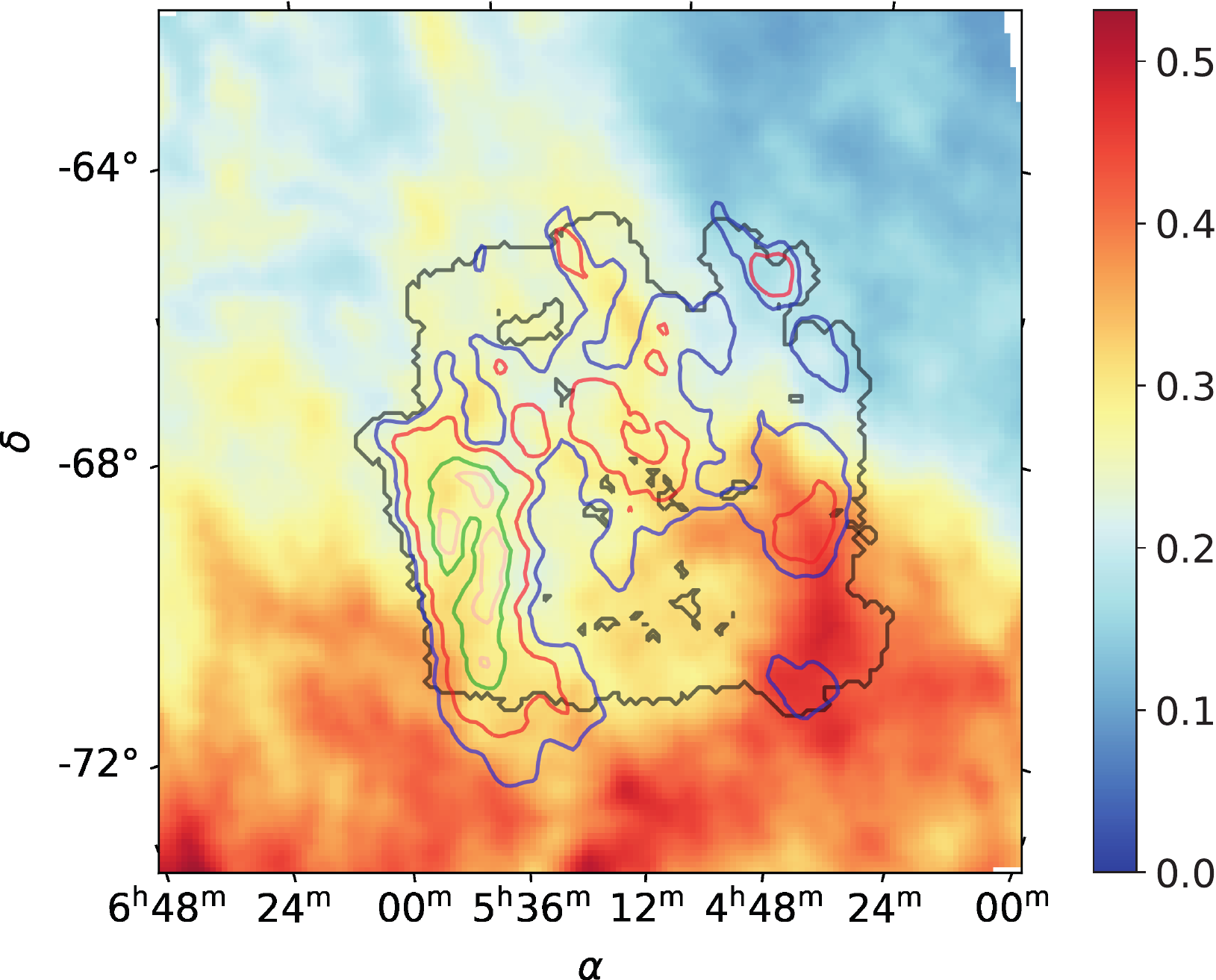} 
    \caption{Map of the foreground dust emission intensity at 353 GHz. Units are MJy/sr. The black contours corresponds to the mask described in Sect.~\ref{sec:planck_data}. The contours are the same  HI integrated intensity levels as in Figure~\ref{fig:HItot}.
    }
    \label{fig:int_fg}
    \end{center}
\end{figure}

\begin{figure}
    \begin{center}
    \includegraphics[width = 0.5 \textwidth]{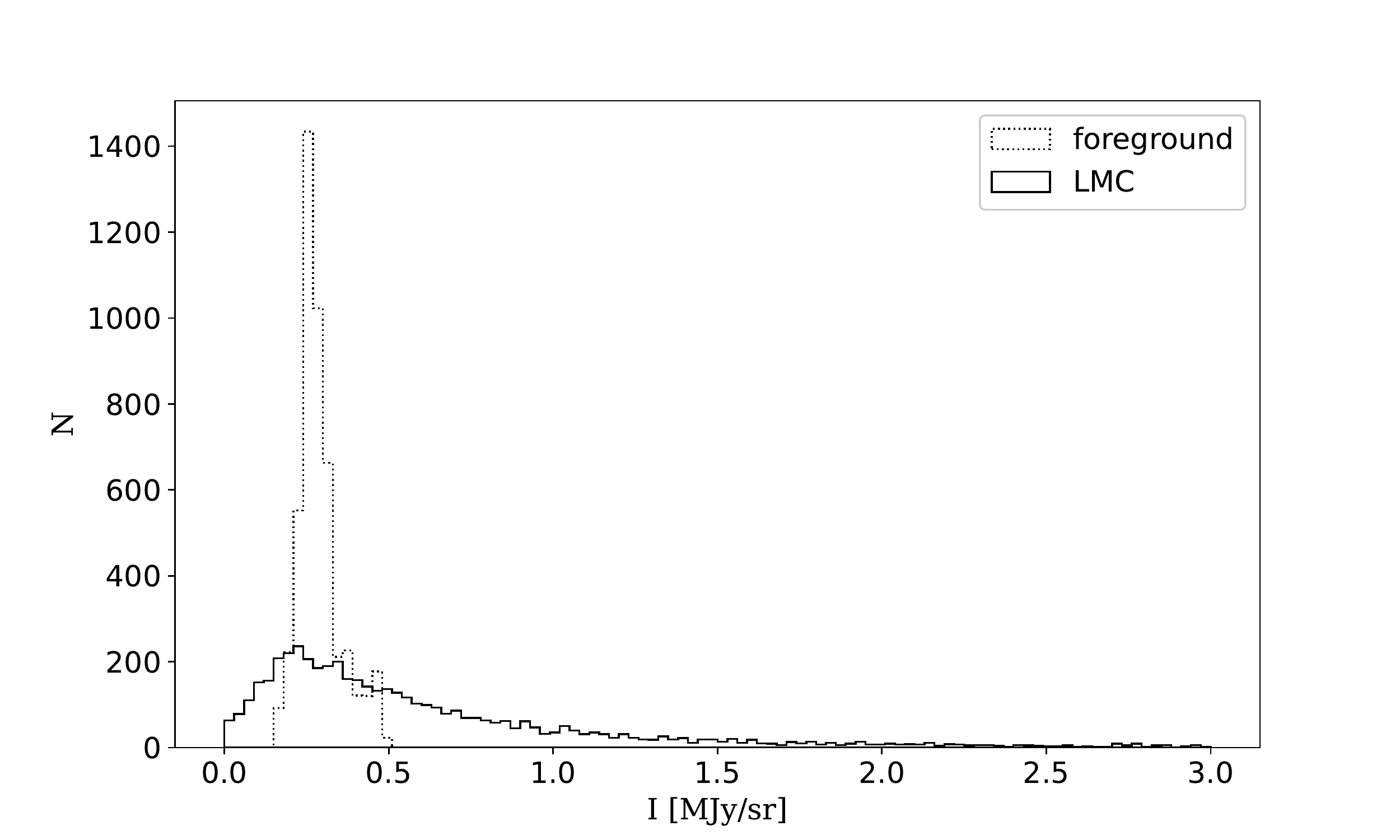} 
    \caption{Histograms of the dust emission intensity taken within the black contour shown in Figure~\ref{fig:int_fg}: for the foreground (dashed curve) and for the LMC after foreground removal (plain curve).}
    \label{fig:i_fg_hist}
    \end{center}
\end{figure}


\adb{We estimate the Galactic foreground intensity $I_{fg}$, assuming that the MW dust emission mostly originates from optically thin atomic gas and that dust properties remain uniform within the chosen field of view. This leads to the expectation that}
\begin{equation}
    I_{fg} = N_H \times \alpha \,\, \label{equ:Iforeground}
\end{equation}
$\alpha$ is a constant and the column density $N_H$ is computed from the HI data using
\begin{equation}
    N_H = 1.823 \times 10^{18} \int T_B dv \,,
\end{equation}
with the integral carried over Galactic velocities.
\adb{We correlated the gas column density from HI data with the original {\planck} intensity at 353 GHz in the outer region and found a slope of} $\alpha = 1.87$ ($\times 10^{21}$). \adb{We used this value to estimate the foreground intensity over the LMC using} Eq.~\ref{equ:Iforeground}.
The resulting foreground intensity map is shown on Figure~\ref{fig:int_fg}.

Figure~\ref{fig:i_fg_hist} shows the histogram of the foreground intensity values towards the LMC at 353 GHz as well as the histogram of intensity values of the LMC after foreground removal. The average LMC brightness in this region is 0.68 MJy/sr, to be compared to the average foreground intensity of 0.29 MJy/sr. The foreground emission is therefore generally significantly lower than that of the LMC in total intensity. The resulting intensity of the LMC after foreground removal is shown in Figure~\ref{fig:bobs_bres}, where the outer spiral density structures and central molecular regions can clearly be distinguished.

\subsection{Determination of the foreground polarization fraction}
\label{sec:p}

\noindent The polarization fraction map derived from the {\planck} data at 353 GHz towards the LMC is shown in Figure~\ref{fig:pmap}.
We measured the average value of $p$ around the LMC from this map, excluding the area delimited by the red contour around the LMC, shown on the figure. 
The red contour is built on the basis of a circle of $5^{\circ}$ radius around the center of the LMC. In addition, we excluded areas where SNR($p$) $<\, 3$ to avoid bias.
The resulting average polarization fraction is $p_{fg} = 7.2\%$, which we assume constant over the whole LMC surface to apply to the foreground emission. This value is close to the value derived for the MW from the Planck data, on average, which is around $7\%$ \citep{planck2014-xix}. Although the polarization fraction of the emission of the foreground material probably varies, we believe this approach to be justified as it does not privilege any particular region and the subtraction is done uniformly. This hypothesis is also supported by the globally uniform foreground magnetic field orientation, as derived from the gradients and observed from the foreground stars polarization.

\section{Results and Discussion}
\label{sec:results}

\begin{figure}
    \begin{center}
    \includegraphics[width = 0.45\textwidth]{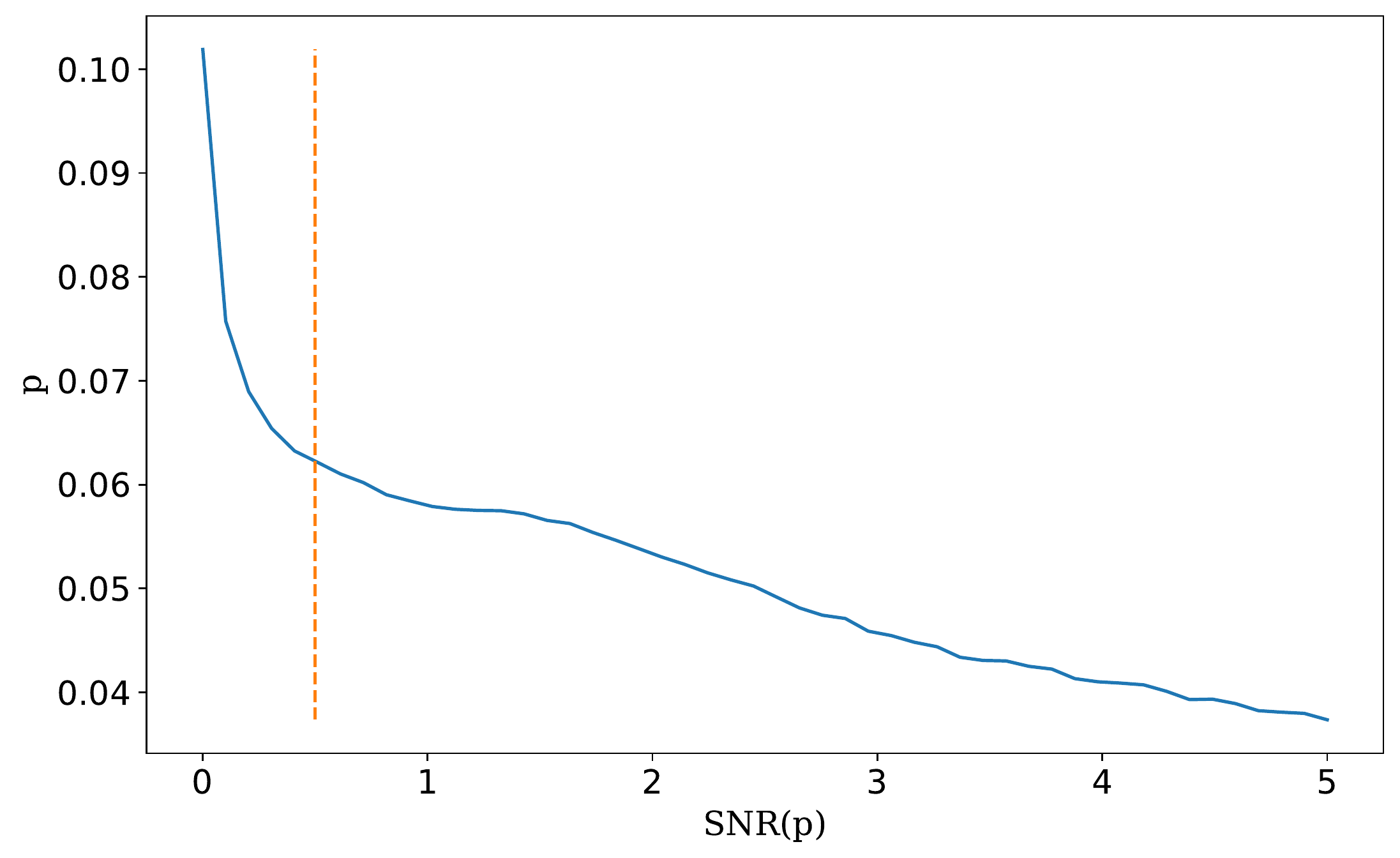}
    \caption{Median polarization fraction in the LMC in the mask defined in Sect.~\ref{sec:results}, after foreground removal as a function of the SNR(p$_{\mathrm{LMC}}$), as estimated according to the procedure described in Appendix~\ref{sec:app}. The red line shows the chosen threshold.}
    \label{fig:snrpnew}
    \end{center}
\end{figure}

\noindent We used the quantities derived in the previous section and Eqs.\,\ref{eq:qu1} and \ref{eq:qu2} to compute maps of the Stokes parameters for the foreground emission $Q_{fg}$ and $U_{fg}$. These maps were then subtracted from the {\planck} 353 GHz data. We also compute the noise variance matrix on the derived Stokes parameters, as detailed in Appendix~\ref{sec:app}. This allowed us to derive the SNR on the derived LMC polarization fraction. \adb{We represent in Figure} \ref{fig:snrpnew} \adb{the variation of the median polarization fraction in the area defined earlier for the LMC as a function of SNR(p$_{\mathrm{LMC}}$). We observe a drastic decrease of p for SNR(p$_{\mathrm{LMC}}$) for values lower than $0.5$.  Thus, the new mask is taken above this threshold and is used for the remainder of the analysis.}  

Below we first describe the resulting polarization angle which gives the large-scale structure of the LMC. Second, we discuss the polarization fraction properties.

\begin{figure}
    \begin{center}
    \includegraphics[width = 0.5 \textwidth]{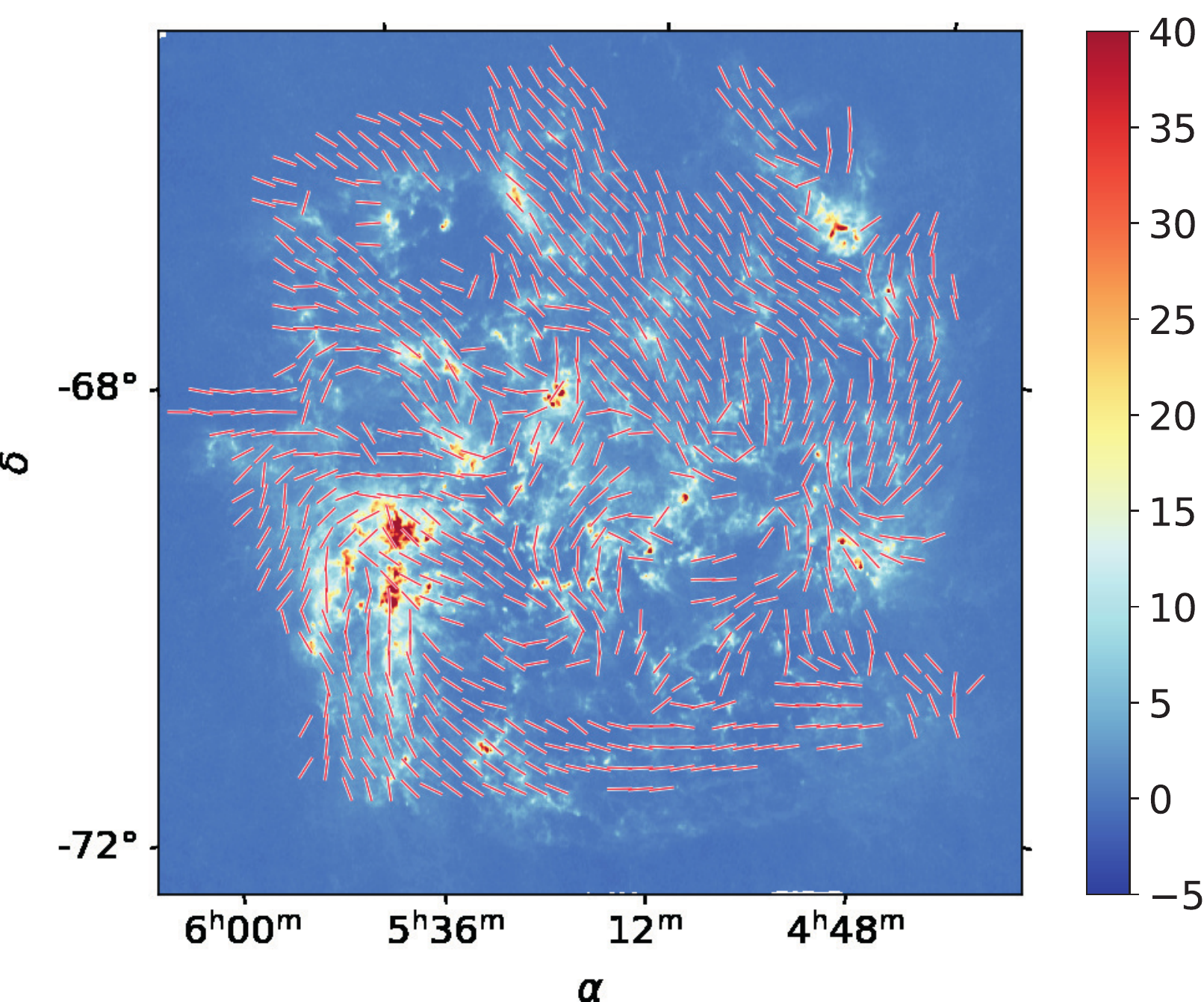}
\caption{Map of the POS direction of the magnetic field derived from the {\planck} data at 335 GHz after foreground removal. The Herschel HERITAGE program's LMC map at 500 microns in MJy/sr is shown in the background for illustration. The directions are not shown for pixels outside the mask defined in Sect.~\ref{sec:results}}
    \label{fig:ilmc_bobs_bres_vec}
    \end{center}
\end{figure}

\begin{figure}
    \begin{center}
    \includegraphics[width = 0.5 \textwidth]{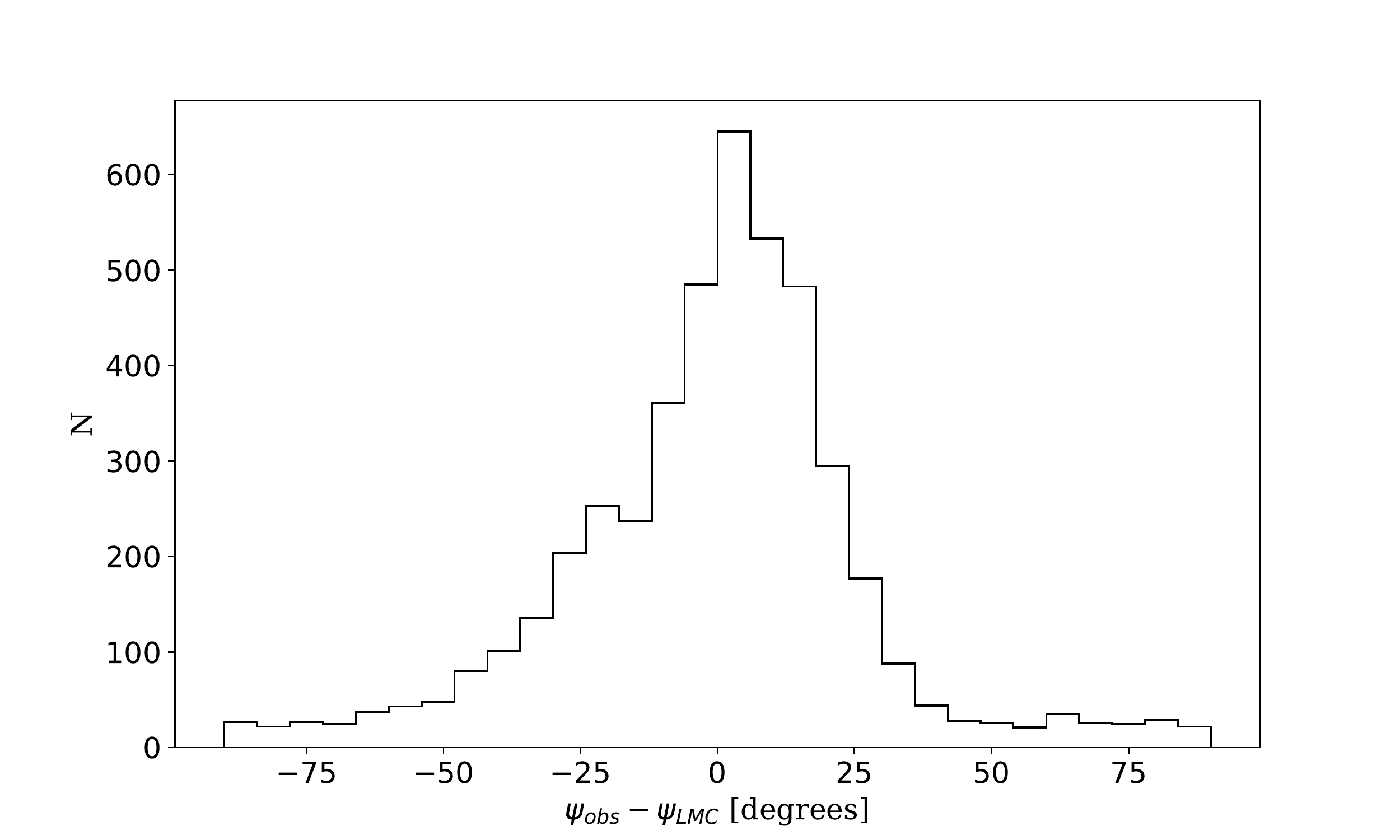}
    \caption{Histogram of the difference between the polarization angles before and after removal of the background. Only pixels inside the mask shown described in Sect.~\ref{sec:results} are taken into account.}
    \label{fig:angdiff_hist}
    \end{center}
\end{figure}

\subsection{Large-scale magnetic field structure}

\begin{figure}
    \begin{center}
    \includegraphics[width = 0.5\textwidth]{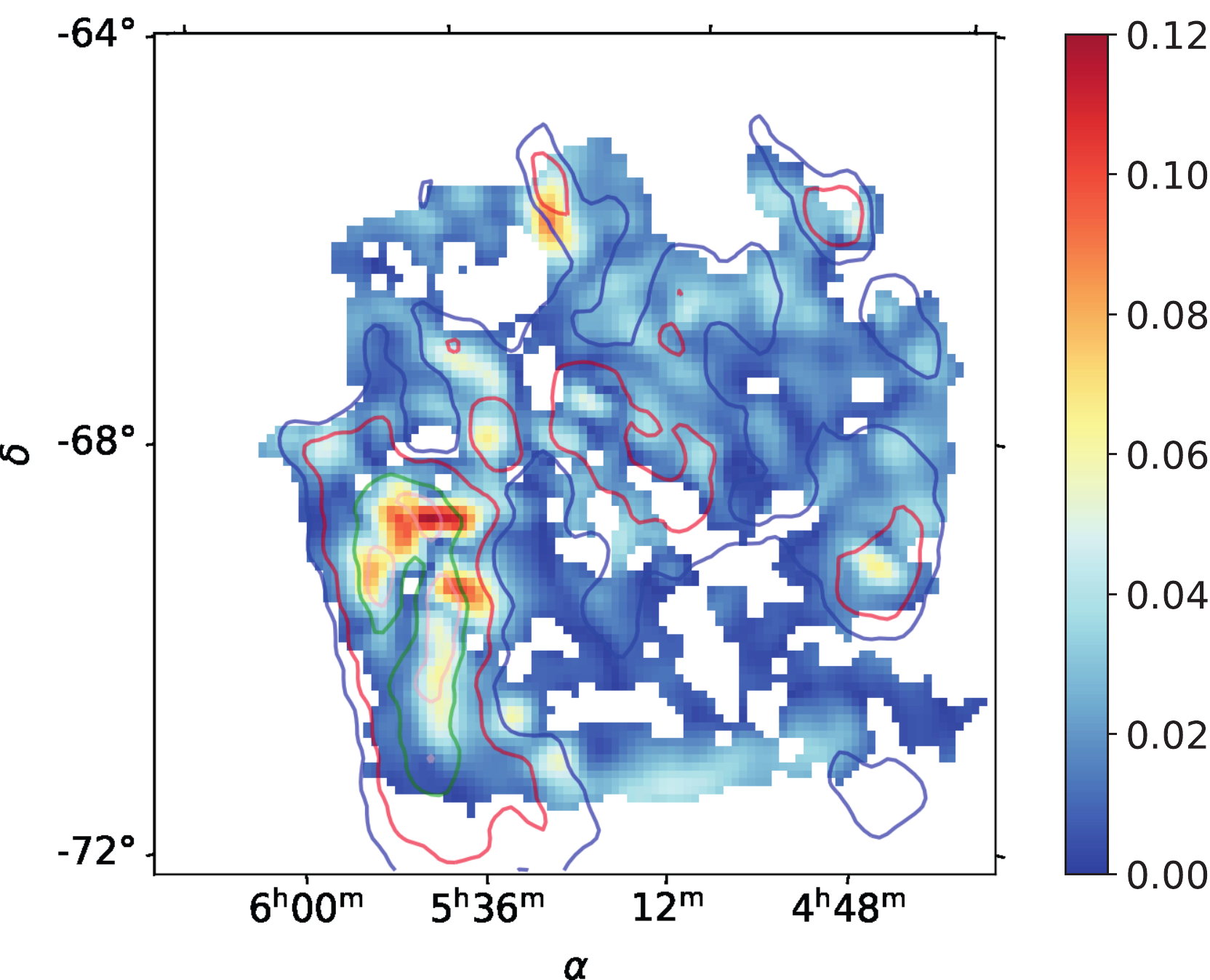}
    \caption{Map of the polarized intensity of the LMC at 353 GHz after foreground removal. Only pixels inside the mask defined in Sect.~\ref{sec:results} are represented. The contours are the same  HI integrated intensity levels as in Figure~\ref{fig:HItot}.}
    \label{fig:pI_Blic}
    \end{center}
\end{figure}

\begin{figure*}
    \begin{center}
    \begin{tabular}{cc}
     \includegraphics[width = 0.495\textwidth]{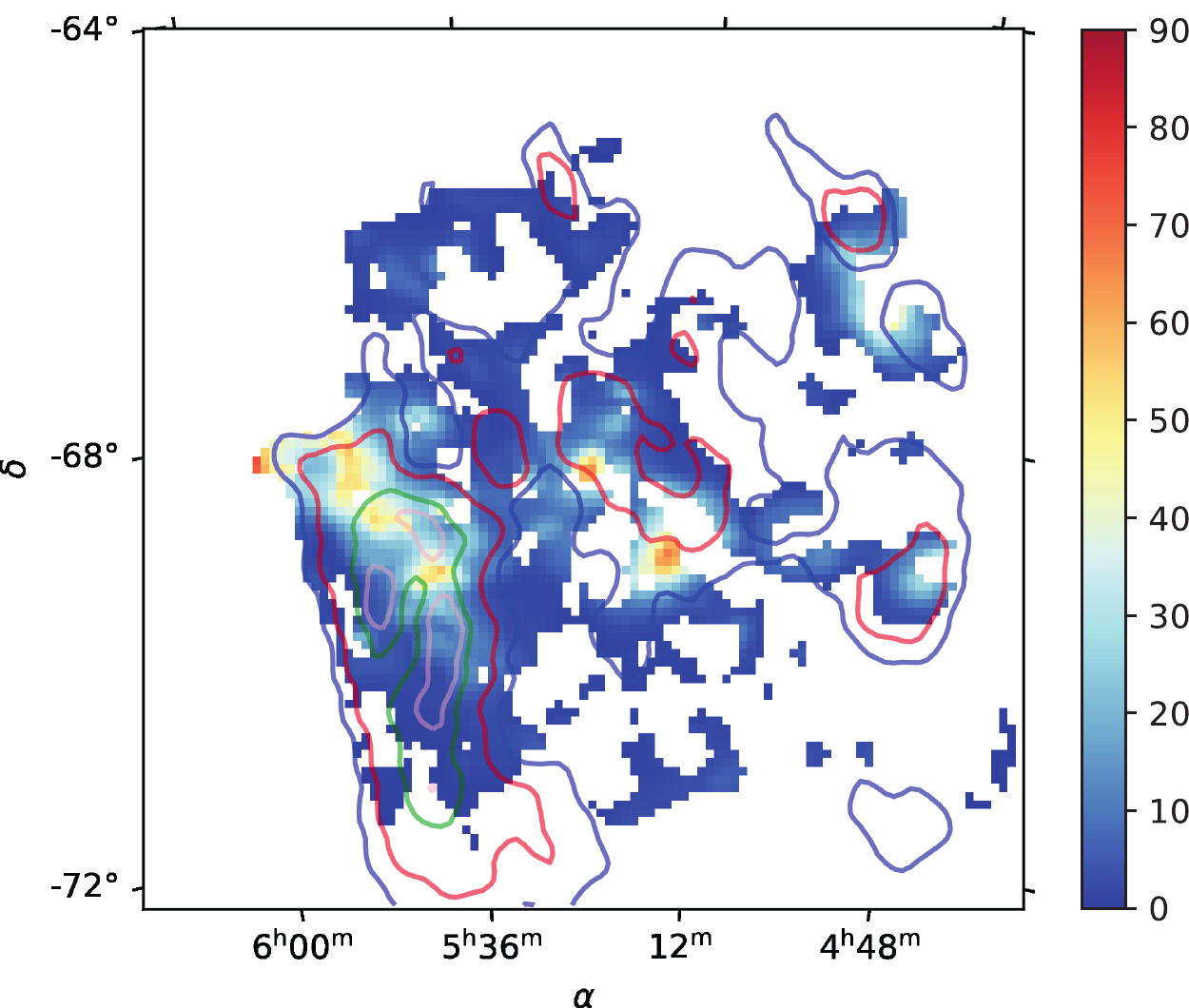} &
     
     \includegraphics[width = 0.495\textwidth]{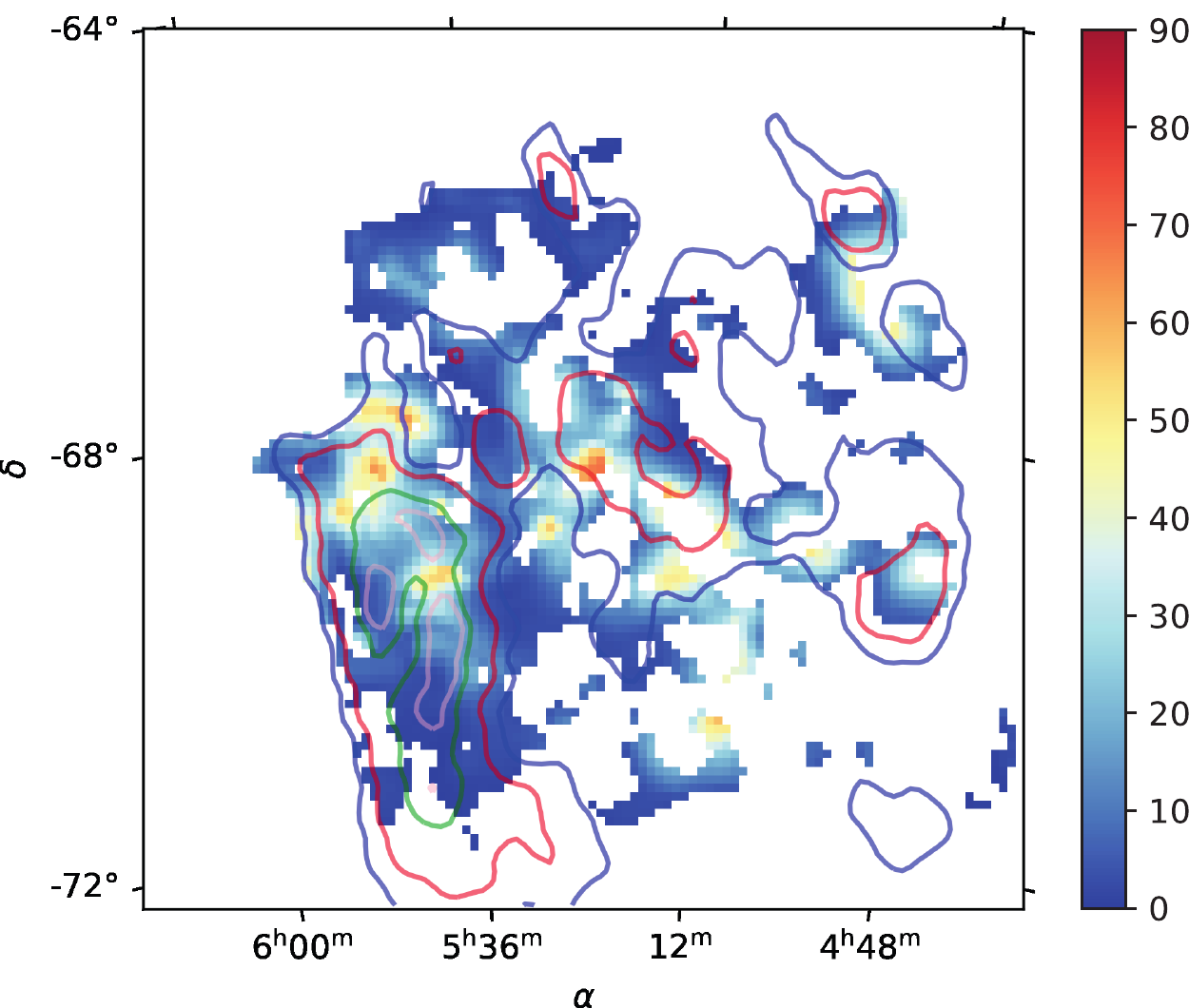}
     \end{tabular}
    \caption{Maps of the polarization angle dispersion function $S$ in degrees, for the initial data (left panel) and the foreground subtracted (right panel) data. Only pixels inside the mask defined in Sect.~\ref{sec:results} are represented. The contours are the same  HI integrated intensity levels as in Figure~\ref{fig:HItot}.}
    \label{fig:smap}
    \end{center}
\end{figure*}

\begin{figure}
   \begin{center}
   \begin{tabular}{cc}
    \includegraphics[height = 0.22\textwidth, trim = {0 0 3cm 0},clip]{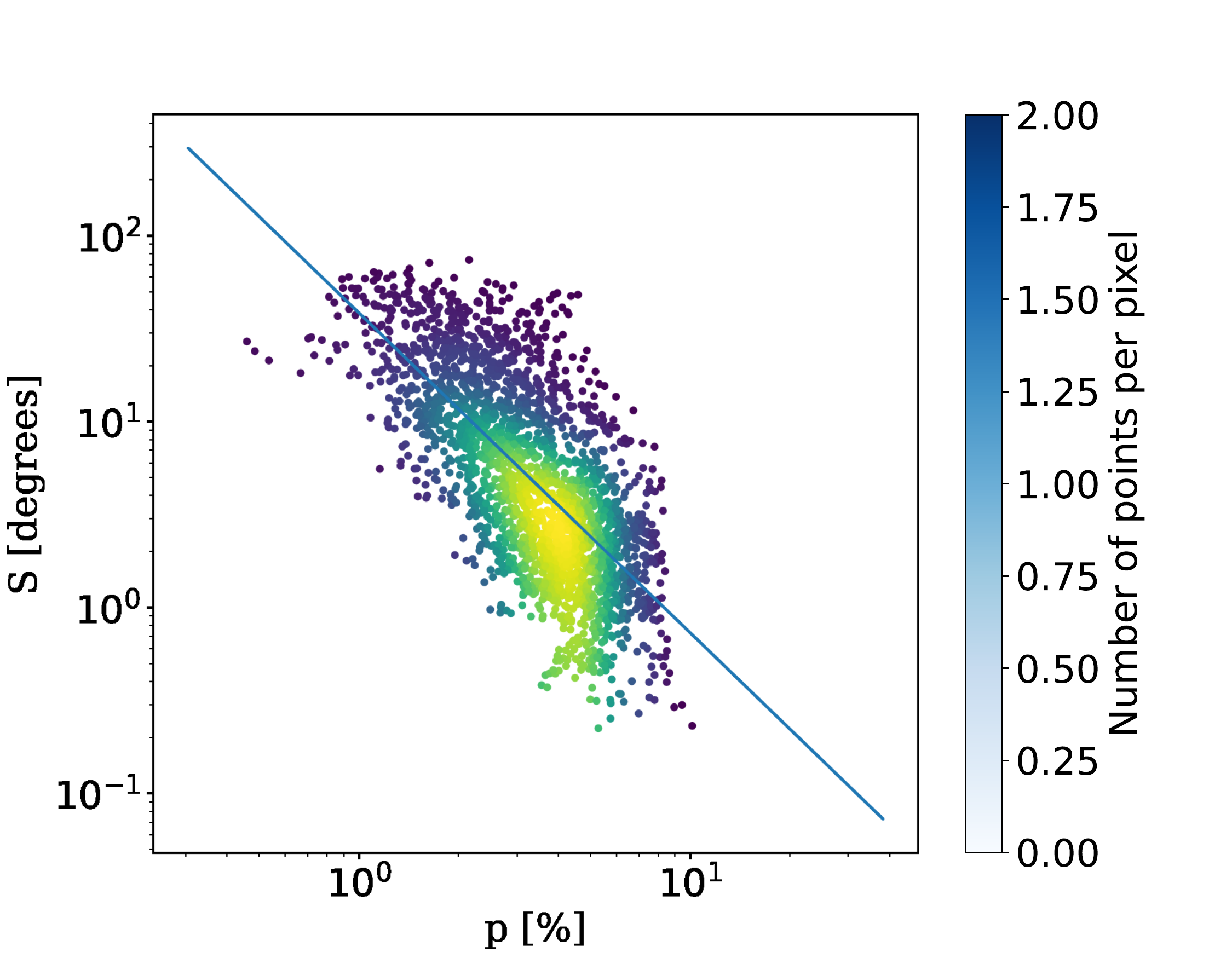}    
    &  \hspace{-0.4 cm}
    \includegraphics[height = 0.22\textwidth, trim = {0 0 0 0},clip]{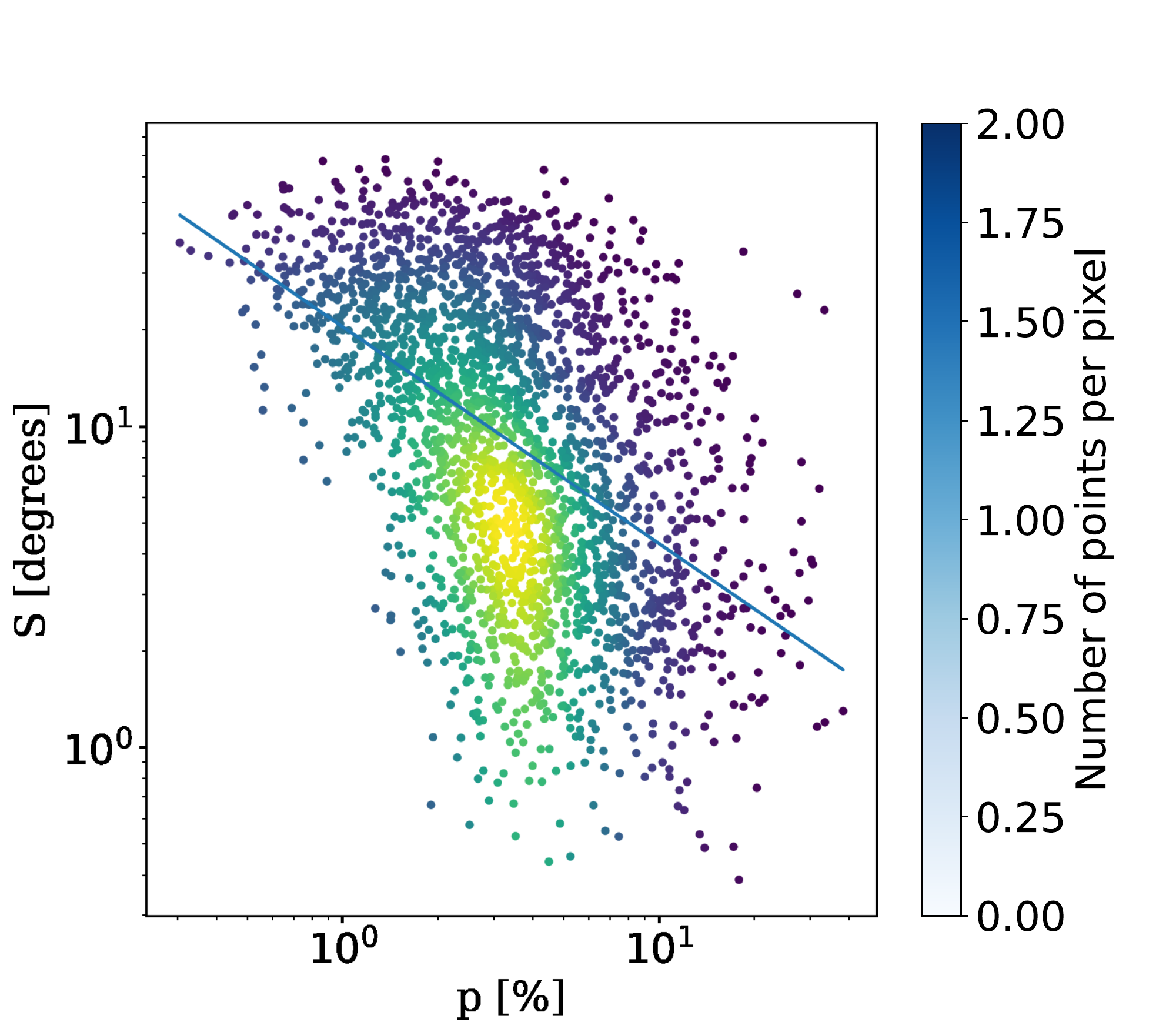}
   \end{tabular}
   \caption{Density scatter plots of $p$ and $S$ in logarithmic scale using the initial polarization signal (left panel) and using the foreground subtracted signal (right panel). The red lines show the best fits obtained using Eq.\ref{eq:log}.}
   \label{fig:p_vs_s}
        \end{center}
\end{figure}

\noindent Figure~\ref{fig:ilmc_bobs_bres_vec} shows the POS magnetic field direction derived from the {\planck} data at 353 GHZ after foreground removal.  
Although the initial and the foreground subtracted magnetic field directions are similar over the brightest LMC regions, as can be seen from Figure~\ref{fig:bobs_bres}, the foreground removal produces a rotation of the magnetic field directions over faint regions the LMC, especially in the north-east quadrant.
The histogram of the angles difference before and after the foreground removal is shown on Figure~\ref{fig:angdiff_hist}. We observe that the angles differ by at most $20^{\circ}$ at the $3\sigma$ level.
The alignment between intensity structures of the LMC and the magnetic field direction appears tighter after foreground removal than in the initial data, in particular, in the southern parts of the LMC.
We also observe that the magnetic field of the LMC after foreground removal tends to align with the outer faint arms. Moreover, we observe an alignment between the faint inner structures and the magnetic field, south-east from the center. 
On the contrary, in a part of the north-western region of the LMC, we still observe the regular north-west to south-east magnetic field orientation. 
It can be due to an incomplete foreground removal. However, such a direction may also be inherent to the LMC's magnetic field. \citet{schmidt1976} retrieves similar polarization direction for stars located in the LMC, after their foreground removal performed for the visible polarization data (seen in the Figure 3 of \citet{schmidt1976}).  

Figure~\ref{fig:pI_Blic} shows the map of the polarized intensity $P = pI = \sqrt{Q^2+U^2}$ after foreground removal.
The figure shows a faint elongated structure in the outskirts of the southern part of the LMC, which is only clearly apparent in the polarized intensity map. It corresponds to Arm S introduced by \citet{staveley-smith2003}. This feature is clearly seen
in the integrated intensity from HI data, as shown by \cite{mao2012}. However, no peculiar structures in synchrotron polarized intensity are observed in that region. The polarized dust intensity highlights regions with intrinsically high polarized dust emission which allow to detect aligned dust in the low density Arm S of the LMC for the first time. There, the magnetic field shows a smooth spiral structure from the eastern  to the southern part of the LMC.

We use the polarization angle dispersion function $\mathit{S}$ as a measure of the regularity of the magnetic field, given by
\begin{equation}
    S =  \sqrt{ \frac{1}{N(l)}  \sum_{i=1}^{N(l)} \left[ \psi(\mathbf{x}) - \psi (\mathbf{x}+\mathbf{l}_i ) \right]^2    } \, ,
\end{equation}
where $\psi(\mathbf{x})$ and $\psi (\mathbf{x}+\mathbf{l}_i )$ are the polarization angles at locations $\mathbf{x}$ and $\mathbf{x}+\mathbf{l}_i$, respectively, and the sum is taken over all pixels contained within an annulus of radius $l = |\mathbf{l}|$ (called the lag) and width $\delta \, l$ from position $\mathbf{x}$. $\mathit{S}$ measures the average amplitude of the local variations of the polarization angle at scale $\mathit{l}$ \citep{serkowski1958,hildebrand2009,alina2016}.

The maps of $\mathit{S}$ at a lag of $16\arcmin$ before and after foreground removal are shown on Figure~\ref{fig:smap}. We restrict the analysis to the data with SNR(S)$> 3$, where the uncertainty is computed using Monte Carlo simulations that take into account the whole noise covariance matrix. The difference between the maps reflects the difference between the initial and foreground corrected magnetic field directions shown on Figure~\ref{fig:ilmc_bobs_bres_vec}. 
The foreground removal reveals noticeable rotation of the magnetic field especially correlated with locations of density structures.
\cite{planck2014-xix} showed an anti-correlation between $p$ and $\mathit{S}$ in the Milky Way which indicated that, globally, on large scale the decrease of polarization is due to the rotation of the magnetic field angle. We represent on Figure~\ref{fig:p_vs_s} the scatter plot of $p$ versus $S$ for the LMC. We observe a clear anti-correlation trend, similar to what has been observed for the Milky Way in general in the {\planck} data \citep{planck2014-xix}. 
Following the approach used in \cite{planck2014-xix}, we fit a straight line to the data in log-log, using
\begin{equation}
\log_{10} (S) = \alpha \times \log_{10} (p) + \beta \, .
\label{eq:log}
\end{equation}
The fit to the original data leads to
$\alpha = - 0.789$ and $\beta = 1.661$
while the same fit on the foreground subtracted data leads to
$\alpha = - 0.504$ and $\beta = 1.287$. The slope of the correlation before the foreground removal is therefore closer to the value derived for the ISM in the Milky Way, $\alpha =-0.834$, in \cite{planck2014-xix} while it is shallower after the foreground removal. 

\subsection{Relative orientation between magnetic field and the LMC structure}
\label{sec:relative_orientation}

\begin{figure}
   \begin{center}
   \includegraphics[width = 0.5 \textwidth]{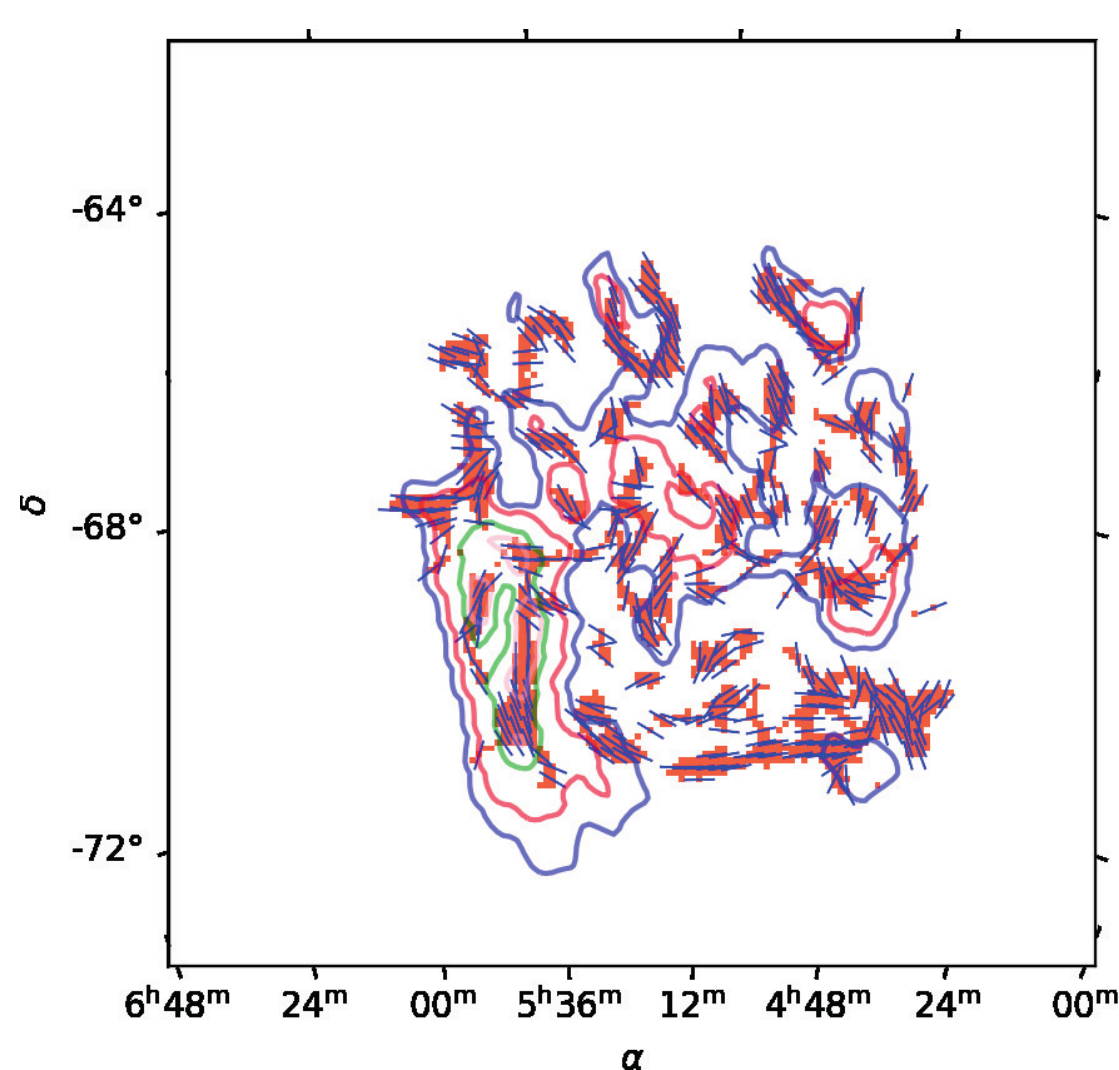}
    \caption{Map of the filamentary structures identified using the RHT algorithm in the {\planck} 353 GHz data, after foreground removal, overlaid with segments showing the POS magnetic field direction in the LMC. Only pixels inside the mask defined in Sect.~\ref{sec:results} are represented. The contours are the same  HI integrated intensity levels as in Figure~\ref{fig:HItot}.}
    \label{fig:rht_map}
    \end{center}
\end{figure}

\begin{figure}
   \begin{center}
   \includegraphics[width = 0.5 \textwidth]{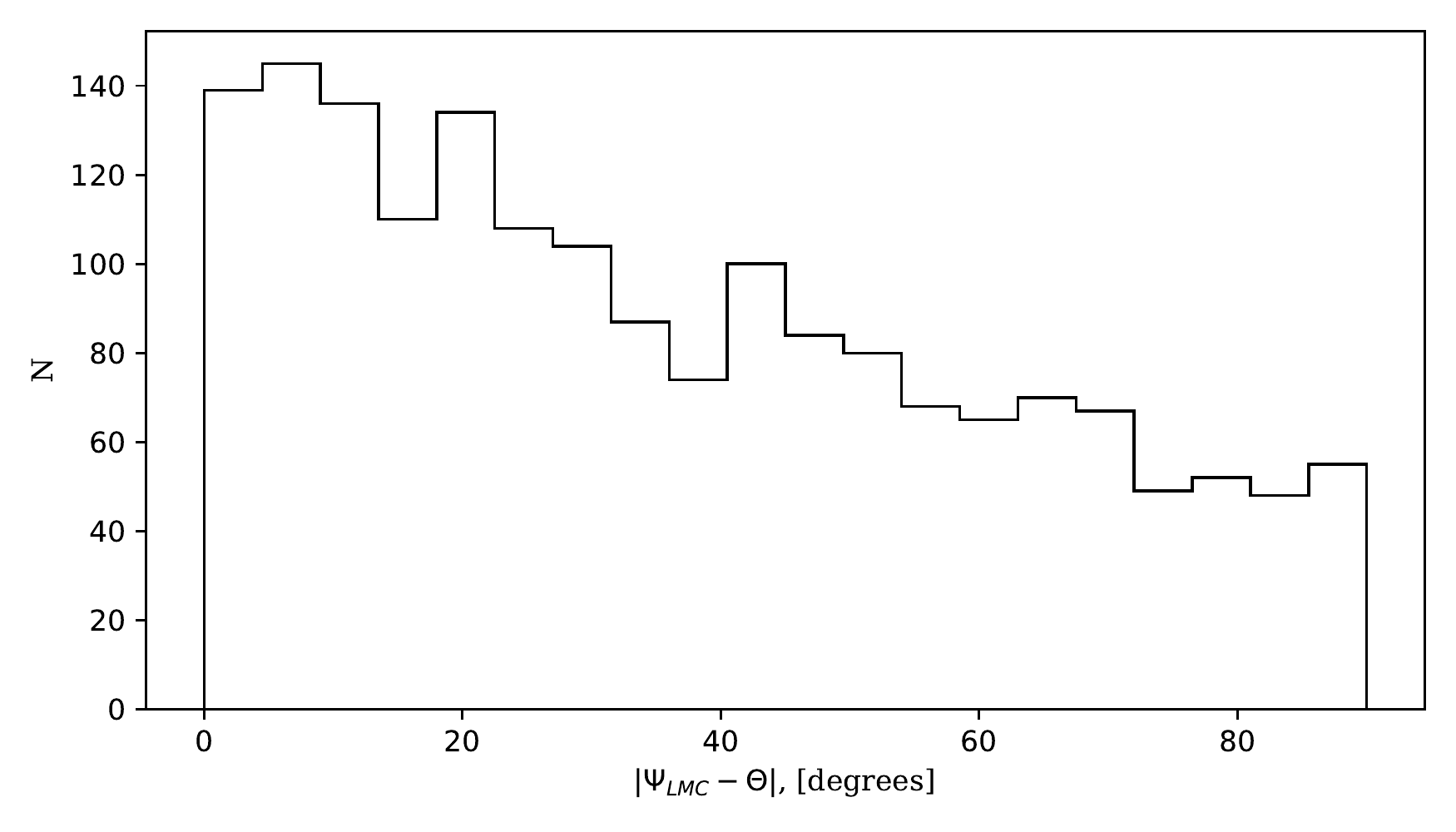}
    \caption{Histogram of the absolute difference between the magnetic field orientation and the orientation of filaments in the LMC after foreground removal, shown in Figure~\ref{fig:rht_map}.}
    \label{fig:relative_orientation}
    \end{center}
\end{figure}
The relative orientation between the magnetic field and the galaxy structure, which is almost seen face-on, provides a hint on the role of the magnetic field in the global dynamics of the LMC.
Although the visual inspection of Figure~\ref{fig:ilmc_bobs_bres_vec} allows us to conclude that in vast regions such as the arms of the galaxy the magnetic field is aligned with density structures, we aim to quantify more precisely their relative orientation. To do so, the RHT algorithm was applied to the {\planck} 353 GHz intensity map, and the best-suited parameters of the kernel's width $16\arcmin$ and length $50\arcmin$ were chosen based on the visual analysis. We represent the corresponding bitmask map on Figure~\ref{fig:rht_map}. The absolute difference between the magnetic field angles and the position angles of the detected filaments in Figure~\ref{fig:relative_orientation} shows that there is a trend for alignment between the magnetic field direction and the density structures in the LMC.

\subsection{Polarization fraction and polarized intensity}
\label{sec:final_parameters}

\begin{figure*}
    \begin{center}
    \begin{tabular}{cc}
    \includegraphics[height
    = 0.4\textwidth]{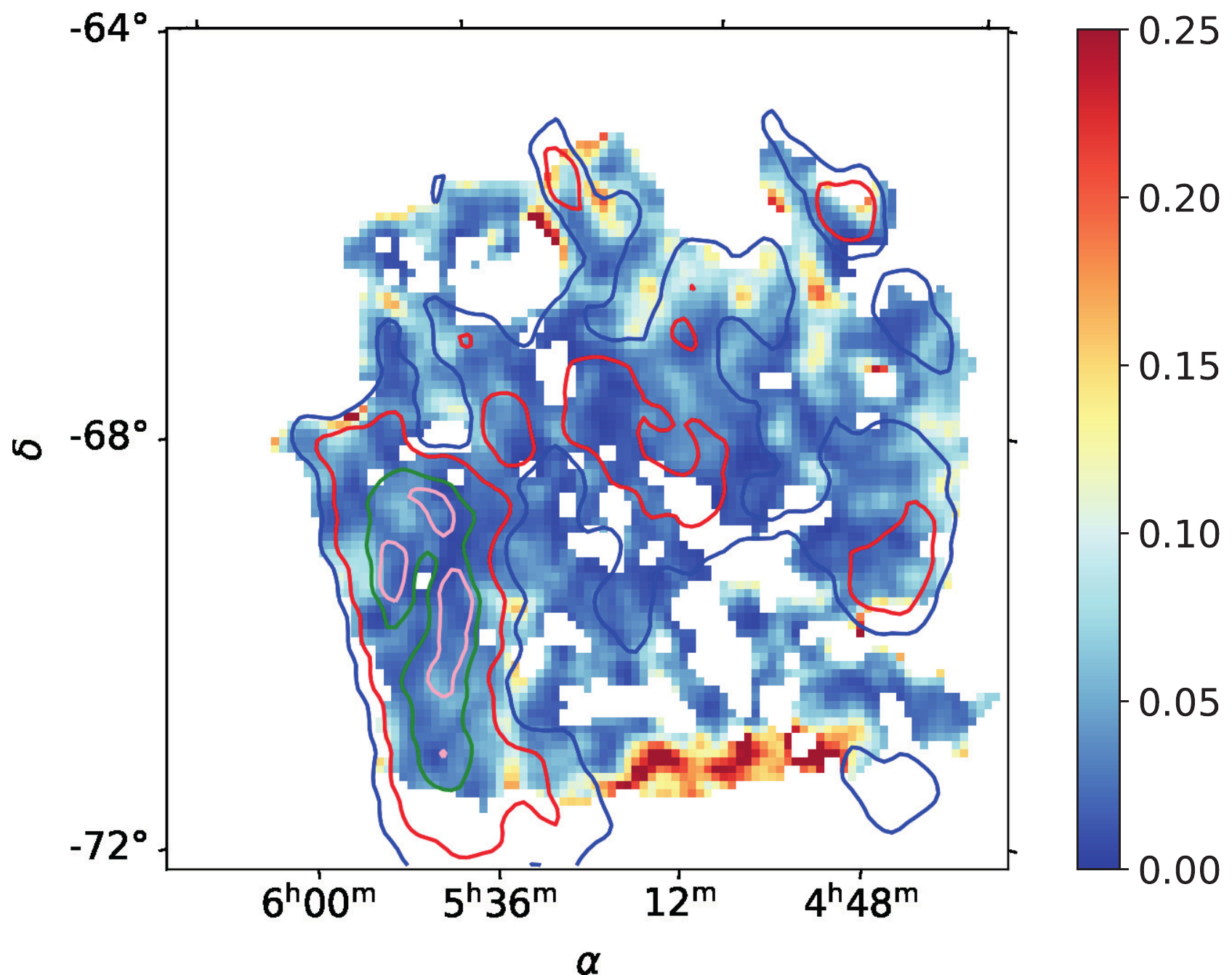}
&
    \includegraphics[height = 0.4\textwidth]{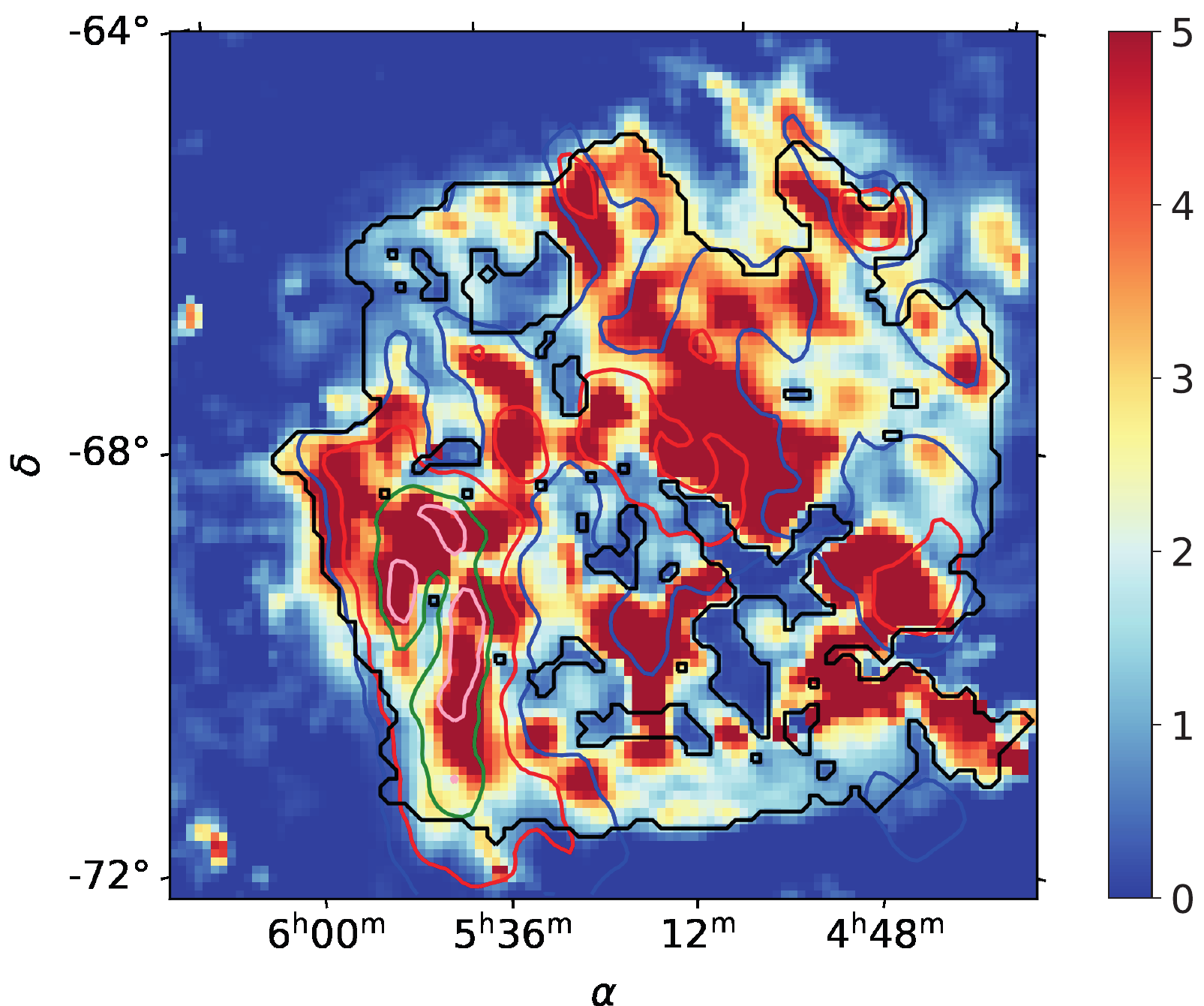}

\end{tabular}
    \caption{Polarization fraction of the LMC, after foreground polarization removal (left panel) and the corresponding SNR (right panel), computed according to the method described in Appendix~\ref{sec:app}. The blue, red, green and pink contours are the same HI integrated intensity levels as in Figure~\ref{fig:HItot}. The black contour delimits SNR($p_{\mathrm{LMC}}$) > 0.5.}
    \label{fig:p}
    \end{center}
\end{figure*}

\begin{figure}
    \begin{center}
    \includegraphics[width = 0.45 \textwidth]{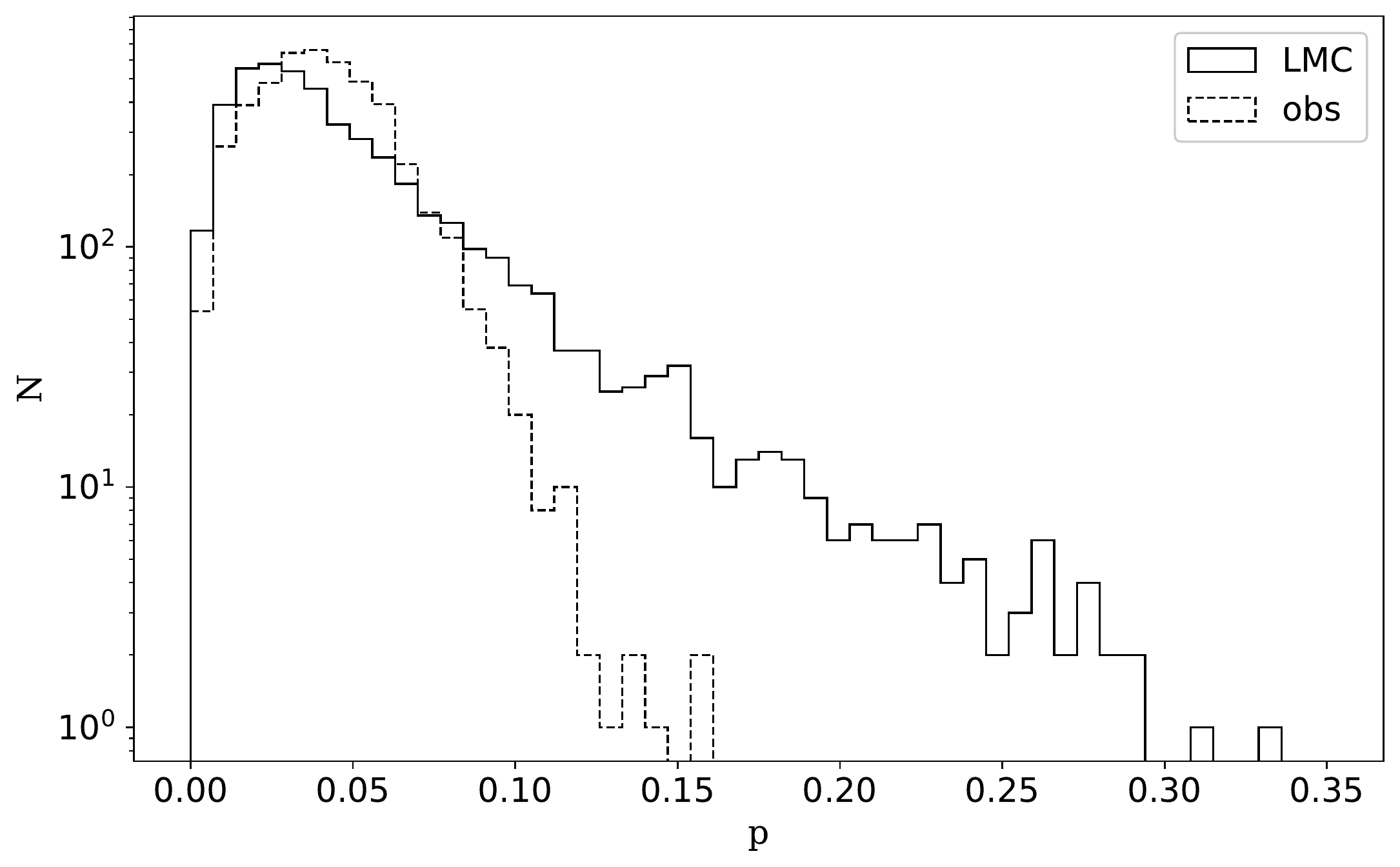}
    \caption{Histograms of polarization fraction in the LMC before (dashed curve) and after (plain curve) foreground removal for pixels enclosed in the mask defined in Sect.~\ref{sec:results}.}
    \label{fig:phist}
    \end{center}
\end{figure}

\begin{figure}
    \begin{center}
    \includegraphics[width = 0.45 \textwidth]{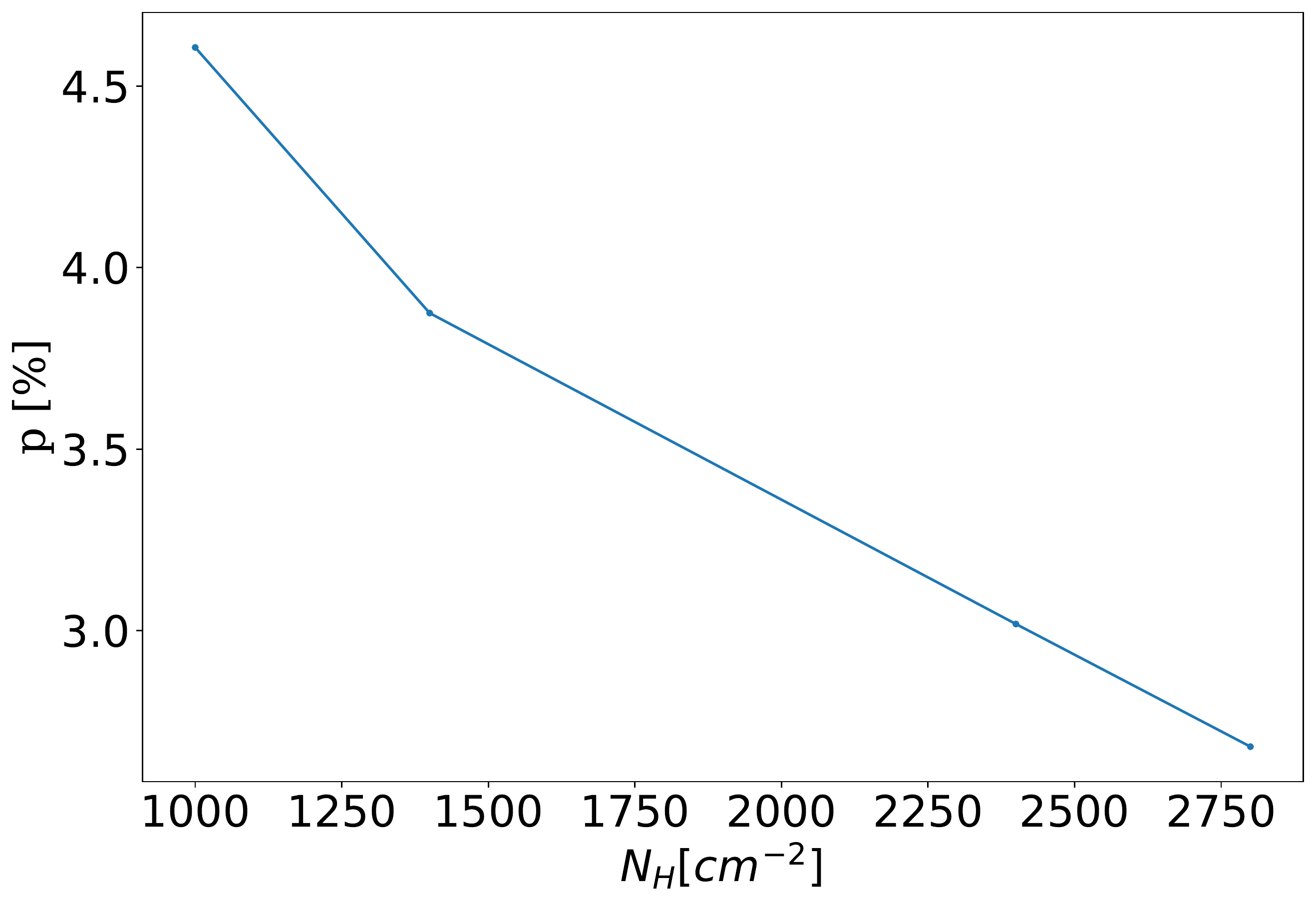}
    \caption{Median value of polarization fraction in the LMC after foreground removal in the regions of increasing gas column densities shown as contours in Figure~\ref{fig:HItot}.}
    \label{fig:pnh}
    \end{center}
\end{figure}

\noindent The map of polarization fraction $p$ after foreground polarization removal is shown in \adb{the left panel of} Figure~\ref{fig:p} \adb{and the corresponding SNR is shown in the right panel.}
Globally, $p$ varies between 2$\%$ and 12$\%$, and in some regions reaches around 20$\%$. \adb{We also observe values that reach around} 25$\%$, \adb{in particular at the location of the foreground structure detected in the HI map.} Although these values may be real, they may be due to noise or to the imperfect foreground subtraction. 
Figure~\ref{fig:phist} shows the histogram of $p$ after foreground removal for the pixels within the mask defined in Sect.\,\ref{sec:p} and Figure~\ref{fig:p}.
We observe that the foreground removal broadens the distribution. 
The median value is $p_{med} =$ {\pobsmed}$\%$ while the original median polarization is $4 \%$. Figure~\ref{fig:pnh} shows how the median value varies with increasing column density. As expected, as the column density increases, corresponding to increasing intensity values, the polarization fraction decreases. 
More details can be found in Table~\ref{tab:p} where we report statistics of the polarization fraction values.
We follow the approach by \citep{planck2018-XII} and show in Table ~\ref{tab:p} the upper 90th, 95th and 99th percentiles. We take the 99th upper percentile  as the conservative estimate of the lower limit of the maximum polarization fraction that is observed in the most favorable conditions, that is, when the magnetic field is oriented in the plane of the sky over most of the LOS. This value is $p_{\mathrm{max}}$ = {\pmaxref}$\%$.
As the LMC is indeed viewed almost face-on, we have here an indication on the maximum polarization observed by {\planck} which is limited by beam dilution and LOS averaging. 
This value is close to the corresponding values derived for the Milky Way of $19.8 \%$ \citep{planck2014-xix} and the value of $22\%$ \citep{planck2018-XII} derived from the Planck data at the resolutions of $60\arcmin$ and $80\arcmin$ respectively. 
We note that the average gas column density of the LMC estimated inside the mask is $1.7 \, 10^{21}$ cm$^{-2}$ which is even higher than the average value of the MW, around $1 \, 10^{21}$ cm$^{-2}$ \citep{planck2014-xix}.
Planck evidenced a systematic decrease of the polarization fraction with column density \citep{planck2014-xix}, and the difference in column density \adb{would lead to lower polarization fractions. However, this is not observed} here for the LMC. 
The difference may indicate that the intrinsic polarization by dust particles is actually higher than the value derived from galactic regions of the solar neighborhood. It could also indicate that the magnetic field in the LMC is more ordered than that of the MW. 

\cite{Vandenbroucke2021} simulated polarimetric observations of external galaxies, including dust properties, alignment, radiative transfer and spiral magnetic field geometries with different inclination angles, based on the MHD structure predicted by the simulations of Milky Way type galaxies. The latter simulations were performed in the frame of the Auriga project \citep{grand2017}. The model, when observed from within the simulated galaxy, was constrained using the {\planck} data and was used to derive linear polarization fractions that could have been observed with the SPICA/B-BOP instrument.
When observed from a distance of 10 Mpc
with the size of 50 kpc, the angular resolution of the SPICA/B-BOP instrument ($38"$) corresponds to the {\planck} resolution in the LMC. 
Their derived value of the lower limit for the maximum polarization fraction at this resolution is around $13\%$ and the average value of the polarisation fraction is around $6\%$. As stated by \cite{Vandenbroucke2021}, to determine the maximum polarization fraction of an external galaxy, it should be resolved with at least 1 kpc, and the LMC observations with {\planck} largely satisfy this requirement. The value we derive here for the polarization fraction of the LMC are therefore lower than the expectations for the average value and significantly higher for the maximum value.

In order to derive an approximate predicted value of polarization fraction that would be observed if the LMC was an unresolved source, we average original $I,\,Q$ and $U$ (Mean Bayesian estimates) inside a circle of a radius of $5^{\circ}$ and obtain $1.1\%$ while the average obtained using the foreground removed parameters, is $3.3\%$.

\begin{table}
\begin{threeparttable}
    \centering
    \begin{tabular}{l|c|c|c}
    \hline \hline
    p [$\%$] &  LMC  &  obs & MW \\
    \hline
     $\mathrm{max(PDF)}$ [$\%$] \dotfill   & \adb{2.5}   & \adb{3.9} & \\
     median $p$ [$\%$]    \dotfill         & {\pobsmed}  & 4.0 & 5.3\tnote{1} \\
     mean p [$\%$] \dotfill                & {\pobsmean} & \adb{4.2} & \\
     upper 90 perc. [$\%$]                 & \adb{9.8}  & \adb{6.7}  & 12.68\tnote{2}\\ 
     upper 95 perc. [$\%$]                 &\adb{13.2}  & \adb{7.8}  & 15.0\tnote{2}\\
     upper 99 perc. [$\%$]                 & {\pobsmax} & \adb{10.0}  & 18.7\tnote{2}
    \end{tabular}
    \caption{Statistics of polarization fraction $p$ in percent after foreground removal, over the pixels enclosed in the black contour shown in Figure~\ref{fig:pmap}, corresponding the description in column 1. 
    These are the polarization fraction at the maximum of the PDF, the median and mean values and the upper 90, 95 and 99 percentiles. Column 2 gives the quantities as measured in the foreground subtracted maps. Column 3 gives the corresponding values in the original {\planck} data. 
    Column 4 gives the corresponding values for diffuse Milky Way ISM as published in the {\planck} papers.}
    \label{tab:p}
\begin{tablenotes}\footnotesize
\item [1] Data taken from \cite{planck2014-xix}.
\item [2] Data taken from \cite{planck2018-XII}, where resolution $80\arcmin $ was employed. Fiducial offset values are taken, with the bias removed as indicated in \citet{planck2018-XII}.
\end{tablenotes}
\end{threeparttable}
\end{table}


\section{Summary and perspectives}
\label{sec:summary}

The aim of this work was to isolate the polarization signal from the Large Magellanic Cloud (LMC) in the {\planck} polarization data at 353 GHz.
The main task consisted in estimating the foreground polarization signal from the Milky Way and subtracting it. This required an estimating the foreground total intensity, polarization direction and polarization fraction.

We tested two techniques to determine the foreground magnetic field direction using HI data. The first technique is based on magnetically-aligned filaments paradigm \citep{clark2019}. It is based on observations in diffuse media density filaments are aligned with plane-of-the-sky (POS) magnetic field, while in dense medium their relative orientation is mainly perpendicular \citep{clark2014,planck2015-XXXV,alina2019,xu2019}. The interpretation of the striations observed in the velocity channel maps according to \citep{clark2019} is due to cold HI filaments aligned by magnetic field. 

The second technique is based on MHD-theory, according to which the turbulent velocities induce the striations in velocity channel maps even in the absence of the density filaments. These fluctuations reflect the gradients of velocity amplitude which are perpendicular to the local magnetic field. Therefore the gradients of intensities in channel maps, similar to polarization, represent the underlying magnetic field direction \citep{lazarian2018vgt}. 

We compared results from the two techniques to the starlight extinction polarization data, for which we updated distance estimates using the Gaia data. This allowed us to choose the most distant stars that trace foreground ISM volumes close to what is traced by sub-millimeter emission in polarization.
A practical advantage of these techniques is that they provide a map, while starlight measurements are sparse, which complicates further pixel-to-pixel analysis.

The comparison showed a better agreement for the velocity gradients technique, which we adopted for the rest of the analysis.
The correspondence between the magnetic field direction inferred from velocity gradients and starlight polarization makes us confident that the derived pattern indeed represents the magnetic field structure.

We estimated the foreground intensity using properties of Galactic dust in the vicinity to the line-of-sight to the LMC. We assume that Galactic dust properties in the foreground towards the LMC match the average properties in \adb{our Galaxy around the LMC}. We derived the \adb{relationship between} dust optical depth and gas column density \adb{empirically from the maps of dust emission and}  HI data.
As for the foreground polarization fraction, we assumed it to be constant across the LMC and equal to the average value observed around the LMC.

We estimated and removed foreground contribution from polarization signal in the {\planck} data and derived polarization parameters of the LMC.  Although the observed {\planck} signal is  dominated by the LMC, the foreground removal allows us to derive global characteristics of magnetic field structure and dust properties of the LMC and compare them to the Milky Way.
We note that there are caveats to the method that we propose. First, the VGT technique requires sub-block averaging which provides the coarse grading of magnetic field maps as compared to the resolution of the original HI maps. Second, for the determination of the foreground Stokes $Q$ and $U$ parameters, we assume the polarization fraction of the Milky Way's dust in the direction of the LMC to be uniform, which in fact can vary. However, we do not expect large variations of $p$ because of the slight variations of the angular dispersion function due to the overall uniform magnetic field direction.  Nevertheless, our method allows us to estimate the foreground polarization in a conservative manner.

We also stress that we focus here on large-scale LMC properties, due to the low resolution of data used.
We observe that magnetic field and density structure elongation directions follow each other closer after foreground removal.
In particular, magnetic field and the south-eastern spiral arm (arm S) are well aligned. We also observe a closer alignment of the LMC structures with the magnetic field in the eastern and western outer arms, but also in the inner regions. 
However, in \adb{a fraction of the observed galaxy, that is the central part of the northern} region, the large-scale magnetic field direction persists after foreground removal and is oriented north-east to south-west.
The polarization angle dispersion function shows an anti-correlation with the polarization fraction as it is observed in the Milky Way, with a shallower slope, which indicates that depolarization is mostly due to the magnetic field rotation inside the beam and along the LOS.

A conservative estimate of maximum observed polarization fraction is {\pmaxref}$\%$ which is similar to the value derived for the Milky Way.
We obtained a median value of $p$ of {\pobsmed}$\%$ and an average of {\pobsmean}$\%$. We use results from recent simulations by \cite{Vandenbroucke2021} to show that this is lower than the expected value for a galaxy similar to ours observed faced-on at the distance of the LMC.  The maximum polarization value we observe is however significantly higher than the expected value from those simulations ($13\%$).
The difference could originate from a more ordered large-scale magnetic field structure or from more efficient dust alignment in the LMC compared to our Galaxy. Both alternatives appear plausible. The difference in the magnetic field structure could arise from the difference in the star formation in the LMC and the MW while the difference in the dust alignment efficiency may arise from the different dust composition that have been evidenced by observations of the dust emission \citep{clayton1985,Matsuura2009,gordon2014,Planck2011_LMC_SMC,devis2017}.
The relative role of these factors requires further investigation, which is out of the scope of this paper.

\appendix

\section*{Acknowledgements}
This work was supported by the Science Committee of the Ministry of Education and Science of the Republic of Kazakhstan, in the frame of the project "Study of the polarised emission of the interstellar medium of the Magellanic Clouds using big data analysis and machine learning", Grant No. AP08855858. DA and MI acknowledge the Nazarbayev University Faculty Development Competitive Research Grant Programme No110119FD4503. AL and KHY acknowledge the support of NASA ATP2303
80NSSC20K0542 and NASA TCAN 144AAG1967 as well as NSF AST  1715754 and 1816234. \\ We acknowledge the use of data provided by the Centre d'Analyse de Données Etendues (CADE), a service of IRAP-UPS/CNRS (\url{http://cade.irap.omp.eu}). \\ We thank the anonymous reviewer for the time and effort invested into the review of the manuscript and for the valuable comments and suggestions.

\section*{Data availability}
Upon publication, maps of the foreground-subtracted $I,\,Q$ and $U$ at 353 GHz will be made available via the CADE\footnote{\url{http://cade.irap.omp.eu}} database.

\bibliographystyle{mnras}
\bibliography{main}

\section{Additional figures}

\begin{figure*}
    \begin{center}
    \begin{tabular}{ccc}
    \includegraphics[width = 0.345 \textwidth]{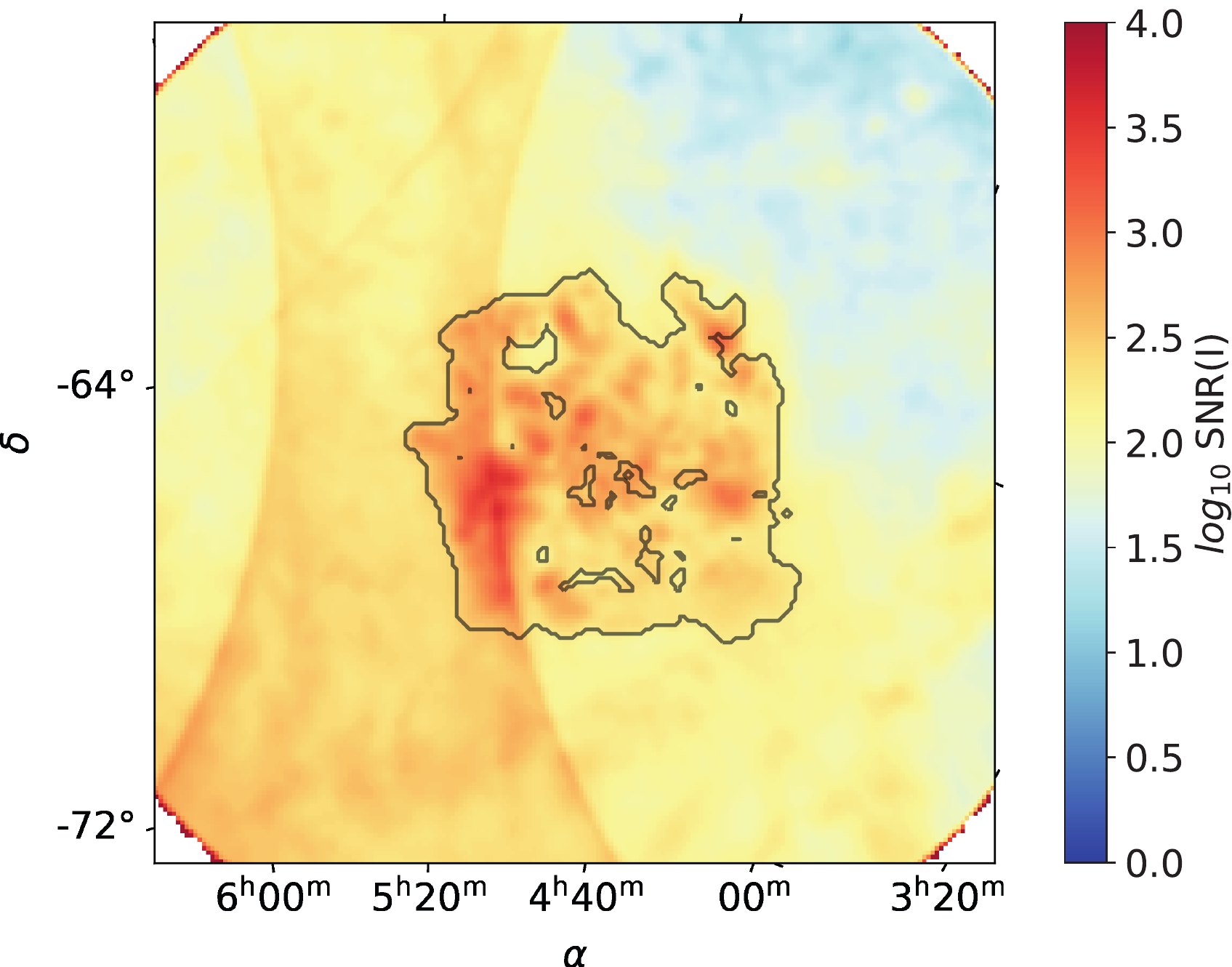}
    &
    \includegraphics[width = 0.33 \textwidth]{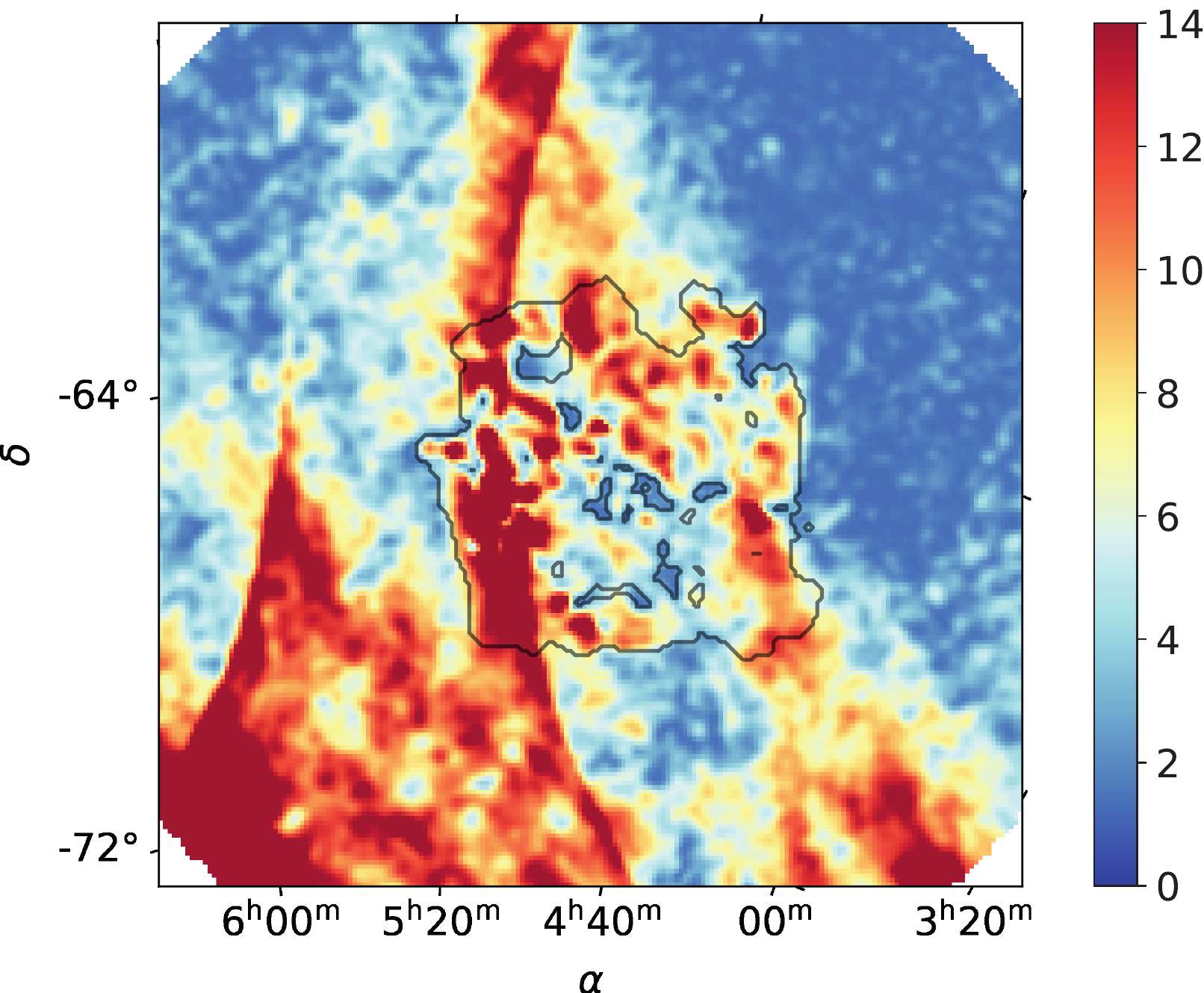}
    &
    \includegraphics[width = 0.33 \textwidth]{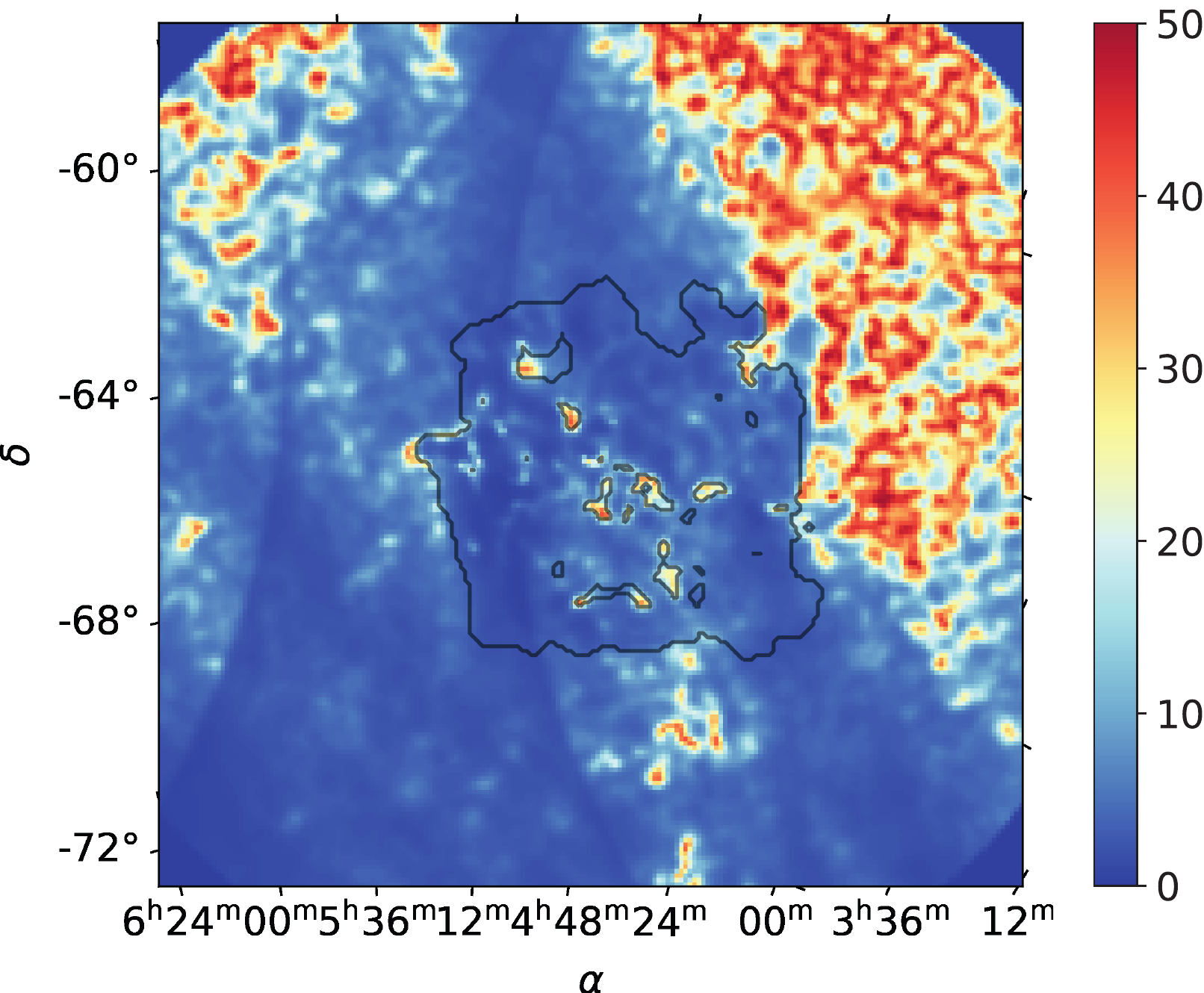}
    \end{tabular}   
    \caption{\textbf{Left:} signal-to-noise ratio of intensity, $I/\sigII$, in log$_{10}$ scale. \textbf{Center:} signal-to-noise ratio of polarization fraction, $p/\sigma_{p}$, where $\sigma_{p}$ is obtained from Bayesian analysis. \textbf{Right:} uncertainty of the polarization angle, $\sigma_{\psi}$, obtained from Bayesian analysis.}
    \label{fig:snrI}
    \end{center}
\end{figure*}

\begin{figure}
    \begin{center}
    \includegraphics[width = 0.5 \textwidth]{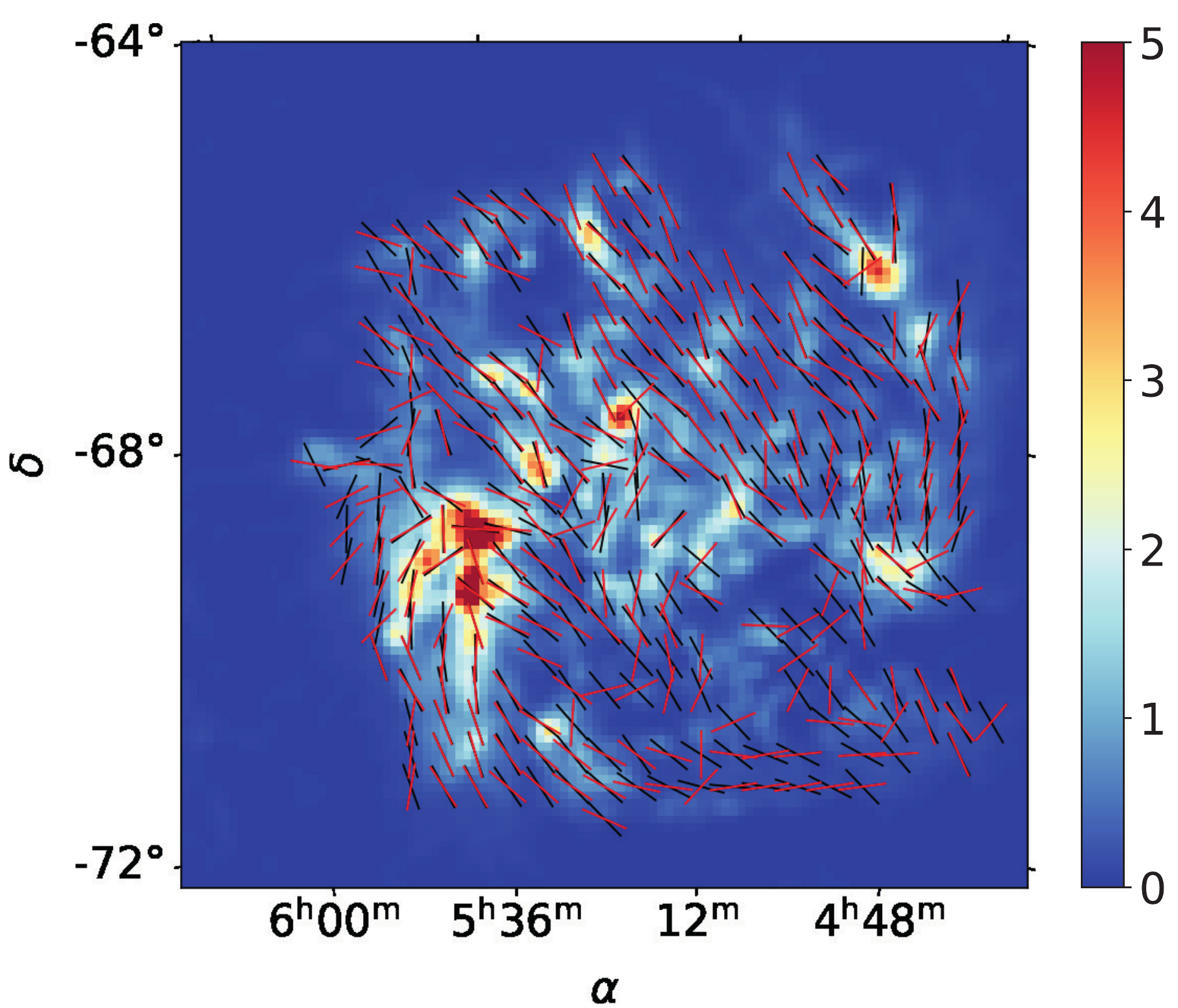}    
\caption{Map of the LMC 353 GHz intensity after foreground removal overlaid with segments showing the direction of the magnetic field derived from the {\planck} data at 335 GHz. The black and red segments show the geometry of the field before and after foreground removal respectively. The directions are not shown for pixels outside the mask defined in Sect.~\ref{sec:results}.}
\label{fig:bobs_bres}   
    \end{center}
    
\end{figure}

Figure~\ref{fig:snrI} shows, from left to right, the maps of the signal-to-noise ratios of intensity ($I/\sigma_{II}$) and polarization fraction ($p/\sigma_{p}$) and the uncertainty of the polarization angle $\sigma_{\psi}$. There, $\sigII$ is given from the original smoothed {\planck} data while $\sigma_{p}$ and $\sigma_{\psi}$ are obtained using Bayesian analysis. 
Figure~\ref{fig:bobs_bres} shows the comparison between magnetic field of the LMC in the original {\planck} data and after foreground removal, represented by black and red segments respectively.

\section{Estimation of the noise covariance parameters after foreground removal}
\label{sec:app}
Below we describe the method we used to estimate the covariances between the Stokes I, Q, and U parameters after the foreground removal.
First, we estimate the covariances in the region around the LMC, outside of the red contour in Figure~\ref{fig:pmap}: $C'_{II}, C'_{QQ}, C'_{UU}, C'_{IQ}, C_{IU}, C_{QU}$, where $C$ denotes the covariance, classically noted as $\sigma^2$, to simplify the writing. Second, we estimate the average intensity in the same region: $I'$. Third, we estimate the variances of the LMC foreground using the following relationship:
\begin{equation}
    C^{fg}_{XY}(i,j) = C'_{XY} \frac{I_{fg}(i,j)}{I'} \, ,
\end{equation}
where $X,Y$ are some Stokes parameters (I, Q or U), $(i,j)$ is a 2D position of a pixel, and $I_{fg}(i,j)$ is the foreground intensity at each pixel estimated in Sect.~\ref{sec:intensity}.
Then, we add them to the initial covariances to obtain the covariances of the parameters after the foreground removal:
\begin{equation}
    C^{LMC}_{XY}(i,j) = C^{fg}_{XY}(i,j) + t \sigma^2_{XY}(i,j)  \, ,
\end{equation}
where $t$ is the factor that takes into account the dfference in the pixels size
in HEALPix and WCS formats. Finally, the variance on the obtained polarization fraction of the LMC is computed using the classical approach \citep{wardle1974} extended to include non-diagonal terms of the covariance matrix \citep{Montier2}:
\begin{eqnarray}
    \sigma^2_p = \dfrac{1}{p^{2}I^{4}} \big(Q^{2}\sigma_{Q}^{2} &+&  U^{2}\sigma_{U}^{2} + p^{4}I^{2}\sigma_{I}^{2} +2QU\sigma_{QU} \nonumber \\
    & - & 2QIp^{2}\sigma_{QI} - 2UIp^{2}\sigma_{UI} \big) \, .
\end{eqnarray}

\end{document}